\documentclass[twocolumn,twocolappendix,trackchanges]{aastex631}

\usepackage{bm}
\usepackage{amsmath}

\accepted{\today}

\submitjournal{ApJ}

\begin{document}

\title{Unveiling the Dynamics of Dense Cores in Cluster-Forming Clumps: A 3D MHD Simulation Study of Angular Momentum and Magnetic Field Properties}

\author[0000-0001-5456-4977]{Shinichi. W. Kinoshita}
\affiliation{Department of Astronomy, the University of Tokyo \\
7-3-1 Hongo Bunkyo,113-0033 \\
 Tokyo, Japan}
\affiliation{National Astronomical Observatory of Japan \\
NINS, 2-21-1 Osawa, Mitaka\\
Tokyo 181-8588, Japan}

\author[0000-0001-5431-2294]{Fumitaka Nakamura}
\affiliation{Department of Astronomy, the University of Tokyo \\
7-3-1 Hongo Bunkyo,113-0033 \\
 Tokyo, Japan}
\affiliation{National Astronomical Observatory of Japan \\
NINS, 2-21-1 Osawa, Mitaka\\
Tokyo 181-8588, Japan}
\affiliation{The Graduate University for Advanced Studies (SOKENDAI)\\
2-21-1 Osawa, Mitaka\\
Tokyo 181-0015, Japan}



\begin{abstract}
We conducted isothermal MHD simulations with self-gravity to investigate the properties of dense cores in cluster-forming clumps. Two different setups were explored: a single rotating clump and colliding clumps. We focused on determining the extent to which the formed dense cores inherit the rotation and magnetic field of the parental clump. Our statistical analysis revealed that the alignment between the angular momentum of dense cores, $\bm{L}_{\rm core}$, and the rotational axis of the clump is influenced by the strength of turbulence and the simulation setup. In single rotating clumps, we found that $\bm{L}_{\rm core}$ tends to align with the clump's rotational axis if the initial turbulence is weak. However, in colliding clumps, this alignment does not occur, regardless of the initial turbulence strength. This misalignment in colliding clumps is due to the induced turbulence from the collision and the isotropic gas inflow into dense cores. Our analysis of colliding clumps also revealed that the magnetic field globally bends along the shock-compressed layer, and the mean magnetic field of dense cores, $\bm{B}_{\rm core}$, aligns with it. Both in single rotating clumps and colliding clumps, we found that the angle between $\bm{B}_{\rm core}$ and $\bm{L}_{\rm core}$ is generally random, regardless of the clump properties. We also analyzed the dynamical states of the formed cores and found a higher proportion of unbound cores in colliding clumps. In addition, the contribution of rotational energy was only approximately 5\% of the gravitational energy, regardless of the model parameters for both single and colliding cases. 
\end{abstract}



\section{Introduction} 
\label{sec:intro}

Gravitationally bound dense cores undergo collapse to form protostellar systems, which eventually evolve into stars. The angular momentum of these cores is a crucial factor in the creation of protostellar systems, as it plays a key role in the formation of protoplanetary disks \citep[e.g.,][]{2002A&A...393..927B,2023ApJ...944..222S}.
The interplay between a rotating accretion disk and a magnetic field is responsible for the launching of protostellar outflows \citep[e.g.,][]{Tomisaka_2002,Matsumoto_2004, Banerjee_2006}. 
The angular momentum of the protoplanetary disk is inherited from the dense core, and previous studies have shown that outflows tend to be launched parallel to the disk's angular momentum \citep[e.g.,][]{Tomisaka_2002,Matsumoto_2004,2009A&A...494..147L}.
In addition, the magnetic field within collapsing cores serves as the primary means for the gas to lose angular momentum through magnetic braking, which could inhibit protoplanetary disk formation \citep{Mellon_2008}. Therefore, the initial structures and distributions of angular momentum, magnetic fields, and their relationship within dense cores are critical parameters in protostellar evolution.

Top-down simulations of cluster formation provide a powerful tool for studying the formation and evolution of star clusters \citep[e.g.,][]{2000ApJ...535..887K, 2003MNRAS.339..577B, 2007ApJ...662..395N, 2011MNRAS.414.2511V, 2014ApJ...797...32P}, allowing us to explore the complex interplay between environmental conditions, angular momentum, and magnetic fields in cluster-forming regions. By simulating the collapse of cluster-forming clumps and the subsequent formation of dense cores, these simulations can shed light on the origin and properties of the angular momentum and magnetic fields of cores, as well as their implications for protostellar outflows and disk formation \citep[e.g.,][]{Chen_2018,Kuznetsova_2019, Kuznetsova_2020, 2022ApJ...925...78A, 2023ApJ...943...76M}. For example, \citet{Chen_2018} investigated the properties of dense cores in magnetohydrodynamics (MHD) simulations of large scale converging flows. They suggest that the internal and external magnetic fields are correlated and the angular momentum of cores is acquired from ambient turbulence. \citet{2023ApJ...943...76M} investigated the filament fragmentation process and the properties and evolution of angular momentum in dense cores. These studies highlight the importance of the interplay between dense regions and their environments in determining core properties. 

Recent observations have revealed the velocity fields of massive cluster-forming clumps ($\sim 10^{3} M_{\odot}$), are often complex, exhibiting two or more velocity components when observed with high-density tracers such as $\rm C^{18}O$ and $\rm H^{13}CO^{+}$. These components are sometimes interpreted as evidence of clump-clump collisions \citep[e.g.,][]{Higuchi_2010, Torii_2011}. \citet{Higuchi_2010} detected 13 cluster-forming clumps with $100-1400M_{\odot}$ in $\rm H^{13}CO^{+}$ line emission and found 4 of them have distinct velocity gradients and multiple components with different velocities. They proposed that collisions between clumps could be a potential mechanism for triggering the formation of clusters to reproduce the observed velocity structures of $\rm H^{13}CO^{+}$ clumps. Clump-clump collision can compress the gas at the overlapped areas of the collision, potentially triggering cluster formation. In particular, offset collisions can produce angular momentum in merged clumps, which would be observed as multiple velocity components.

However, it is also possible that the complex velocity structures are due to the systematic internal motions of the clumps, i.e., gravitational contraction of a single clump with rotation.
\citet{2016ApJ...832..205S} investigated the velocity structure of the S235AB clump, a massive cluster-forming clump, in $\rm C^{18}O$. They found a clear velocity gradient with two well-defined peaks around the center in the position-velocity (PV) diagram. They show that a model of an infalling, rotating single clump provides a good fit to the PV diagram. Additionally, by performing statistical analyses of gas kinematics in massive clumps ($\sim 1\times 10^{3}M_{\odot}$), \citet{Shimoikura_2018,Shimoikura_2022} have shown that some molecular clumps exhibit a velocity structure indicating infalling motion with rotation. They concluded that infalling motion with rotation is a common phenomenon during the early stages of cluster formation.

As described, some hypotheses have been proposed to explain the complex velocity components of clumps, suggesting that star clusters may form under the influence of rotation or collisions.
Therefore, it is necessary to investigate the effects of the environment of the clump, such as rotation or collisions, on the properties of protostellar cores. Moreover, distinguishing the origin of the complex velocity structure based solely on current molecular spectroscopy analysis is challenging, and new observational diagnostics are required. 

In this study, we perform a MHD simulation of cluster-forming clumps with adaptive mesh refinement (AMR) by employing the {\tt Enzo} code \citep{2014ApJS..211...19B}. The simulation ingredients include turbulence and gas self-gravity. In our simulation, we investigated the properties of bound cores under different environmental conditions, including single rotating clumps, colliding clumps, and non-rotating/non-colliding clumps. We focused on analyzing the angular momentum and magnetic fields of the identified cores and examined their dynamics. 

Below, Section \ref{sec:method} describes the method of our simulations and analyses. Results are presented in Section \ref{sec:Results}, including analysis of core angular momentum, magnetic fields, and dynamics. In Section \ref{sec:Discussion}, we discuss the implications of our results. We summarize our results in Section \ref{sec:Summary}.

\section{Method} 
\label{sec:method}
\subsection{Numerical Code}
The numerical code is essentially the same as that of \citet{Kinoshita_2022}. We use the numerical code {\tt Enzo}\footnote{http://enzo-project.org (v.2.6)}, a MHD adaptive mesh refinement (AMR) code \citep{2014ApJS..211...19B}.  The ideal MHD equations were solved using a Runge-Kutta second-order-based MUSCL solver utilizing the Dedner MHD solver and hyperbolic divergence cleaning method (\citealp{DEDNER2002645}; \citealp{2008ApJS..176..467W}). The Riemann problem was solved using the Harten-Lax-van Leer  (HLL) method, while the reconstruction method for the MUSCL solver was a piecewise linear model (PLM). The self-gravity of the gas is included in our simulations. 

The numerical domain is set to $L_{\rm box}=2.8~\rm pc$ cubic. We use a root grid of $256^{3}$ with 5 levels of refinement, corresponding to an effective resolution of $8192^{3}$. Our refinement criterion is based on resolving the Jeans length by eight cells to avoid artificial fragmentation \citep{1997ApJ...489L.179T}. Refinement is allowed until the finest resolution reaches $\Delta x_{\min}=L_{\rm box}/8192\simeq 3.4\times 10^{-4} ~\rm pc$,  where the local number density reaches $n_{\rm crit}\simeq 1.2\times 10^{8} ~\rm cm^{-3}$ in some region. We assumed a mean molecular weight $\mu= 2.3$ and an adiabatic index was set to $\gamma= 1.00001$ for an approximate isothermal assumption.

\subsection{Initial Conditions and Parameters}
We choose initial conditions to match properties of observed clumps. As an initial clump, we set a magnetized gas sphere with uniform density $n_{\rm clump}=1.2\times 10^{4}~\rm cm^{-3}$, isothermal sound speed $c_{\rm s}=0.27~\rm km\cdot s^{-1}$, and radii $R_{\rm clump}=0.7~\rm pc$, giving a mass $M_{\rm clump}\sim10^{3}~M_{\odot}$. The clump is embedded within ambient gas of 10 times lower density, $n_{0}=1.2\times 10^{3}~\rm cm^{-3}$. 

The simulation box is initialized with a large-scale uniform magnetic field $\bm{B}_{0}$, which is parallel to the $z$ axis. We explored initial magnetic field strengths of $B_{0}=10~\mu G$ (weak) and $100~\mu G$ (strong). Then, the ratios of magnetic energy $E_{\rm clump,mag}\equiv B_{0}^{2}R_{\rm clump}^{3}/6$ with respect to the gravitational energy $E_{\rm clump,grav}\equiv-3GM_{\rm clump}^2/5R_{\rm clump}$ are $E_{\rm clump,mag}/|E_{\rm clump,grav}|\approx 2.3\times10^{-3}$ and $0.23$, respectively. 

To approximate the velocity and density fluctuations present in observed clumps, turbulent velocities are generated within the clump material at $t=0 ~\rm Myr$. 
This velocity field chosen for our simulations follows a power spectrum of the Larson law $v_k^2 \propto k^{-4}$ \citep{10.1093/mnras/194.4.809}, with a pure solenoidal component
, where $k$ is the wavenumber for an eddy diameter. We limit our $k$-modes to be $4<\frac{k}{\pi /L_{\rm box}}<20$. We select two turbulence Mach number $\mathcal{M}\equiv\sigma_{v}/c_{\rm s}$, $\mathcal{M}=1.5$ (weak) and $\mathcal{M}=5$ (strong), where $\sigma_{v}$ is the velocity dispersion in the clump. From which, the ratios of turbulent energy $E_{\rm clump,turb}\equiv M_{\rm clump}\sigma_{v}^{2}/2$ with respect to the gravitational energy are $E_{\rm clump,turb}/|E_{\rm clump,grav}|\approx 0.02$ and $0.25$, respectively. Turbulent velocity generates some base level of clump angular momentum. The total initial angular velocities purely from turbulence for $\mathcal{M}=1.5$ and $\mathcal{M}=5$ are respectively $\Omega_{\rm turb}\sim 1.0\times 10^{-15}~\mathrm{rad}~\mathrm{s}^{-1}$ and $\sim 3.0\times 10^{-15}~\mathrm{rad}~\mathrm{s}^{-1}$.
Since our initial clumps do not have enough kinetic and magnetic support, they are prone to gravitational collapse at the beginning. 

Previous observations have indicated that clump velocity structures can be attributed to either infall with rotation or clump collision. To investigate these scenarios, we used two different setups in our simulations: `` Rotation  Setup'' where a single clump contracts with rotation, and the `` Collision Setup '' where two clumps collide (see Section \ref{sec:Rotation Setup} and \ref{sec:Collision Setup}).  Table \ref{tab:parameter} lists the models for both setups and illustrates the parameter space explored. Throughout our subsequent discussion, we will refer to the model names as shown in Table \ref{tab:parameter}. 
 
\subsubsection{Rotation Setup}
\label{sec:Rotation Setup}
In the Rotation Setup, besides the initial turbulent velocity field, we add an angular momentum with constant velocity to the entire clump. The rotational angular velocity is $\Omega_{0}=1.0\times 10^{-13} ~\rm rad~s^{-1}$, from which the ratios of rotational energy $E_{\rm clump,rot}\equiv M_{\rm clump}R_{\rm clump}^{2}\Omega_{0}^{2}/5$ with respect to the gravitational energy is $E_{\rm clump,rot}/|E_{\rm clump,grav}|\approx 0.25$. This value is based on some previous observational works \citep[e.g.,][]{Higuchi_2010,Shimoikura_2018}. 
$\Omega_{0}$ is about two orders larger than $\Omega_{\rm turb}$ and dominant for the rotational motion of the entire clump.

To investigate the effects of the initial magnetic field direction, we consider two arrangements with respect to the rotation axis, namely $\theta_{0}=0^{\circ}$ and $45^{\circ}$, where $\theta_{0}$ is the angle between $\bm{B}_{0}$ relative to the $\bm{\Omega}_{0}$. In other words, $\theta_{0}=0^{\circ}$ means that $\bm{\Omega}_{0}$ and $\bm{B}_{0}$ are parallel, and $\theta_{0}=45^{\circ}$ means that the angle between them is $45^{\circ}$. 

The upper row of Figure \ref{fig:initial_cindition} shows the sample map from one of the Rotation Setup model ({\tt Rot-M5-B10P}) showing the simulated gas structure in column density integrated along $z$ axis. Initially, the gas rotates around a constant rotation axis $\bm{\Omega}_{0}$. Gradually, dense structures develop due to turbulence compression and local gravitational collapse, eventually leading to the formation of dense cores.
The red circles in the figure show the positions of identified cores (see Section \ref{method:Measuring Core Properties}), indicating that cores have formed at various locations within the clump. In Appendix \ref{app:Time evolution of column density}, the time evolution of column density, as viewed along the $x$ axis, is shown. We investigated core-to-core separations in Appendix \ref{app:Nearest neighbor core separation}. 

For comparison, we also consider the setup, referred to as "w/o setup", in which single clump contracts without initial angular momentum with $\Omega_{0}$. All conditions in the w/o Setup are the same as in the Rotation setup, except that there is no initial angular momentum with $\Omega_{0}$ in the former.

\subsubsection{Collision Setup}
\label{sec:Collision Setup}
In the Collision Setup, we investigate the collision of two clumps with an initial impact parameter of $b=R_{\rm clump}$.
Turbulent velocities are generated within the two clumps' material at $t=0 ~\rm Myr$, similar to the Rotation and w/o Setups.
However, unlike the Rotation Setup, the initial two clumps are not rotating.
Due to the off-center collision, the shear motion is converted into rotating motion of the compressed dense gas, whose angular momentum axis is roughly perpendicular to the collision axis.
Henceforth, we refer to the axis of rotation generated by the collision as $\bm{\Omega}_{\rm col}$.
The default initial velocity of the clumps is set to be $V_{\rm 0}\approx 2.8~\rm km~s^{-1}$ (fast) by means of relative collision velocity $V_{\rm rel}=2V_{\rm 0}$. For comparison, $V_{\rm 0}\approx 1.4~\rm km~s^{-1}$
(slow) is also explored. From which, the ratios of 
kinetic energy due to the overall motion of the clump, $E_{\rm clump,col}=M_{\rm clump} V_{\rm 0}^{2}/2$ to the gravitational energy are $E_{\rm clump,col}/|E_{\rm clump,grav}|\approx0.25$ and 1.00 respectively. 

We select two arrangement of initial magnetic field $\bm{B}_{0}$ and $\bm{\Omega}_{\rm col}$, $\theta_{0}=0^{\circ}$ and $45^{\circ}$. When $\theta_{0}=0^{\circ}$, it means that the collision axis is perpendicular to $\bm{B}_{0}$ ($\bm{B}_{0} \parallel \bm{\Omega}_{\rm col} $), and when $\theta_{0}=45^{\circ}$, it means that the angle between the collision axis and $\bm{B}_{0}$ is $45^{\circ}$.

\begin{figure*}
\begin{center}
  \includegraphics[width=13cm]{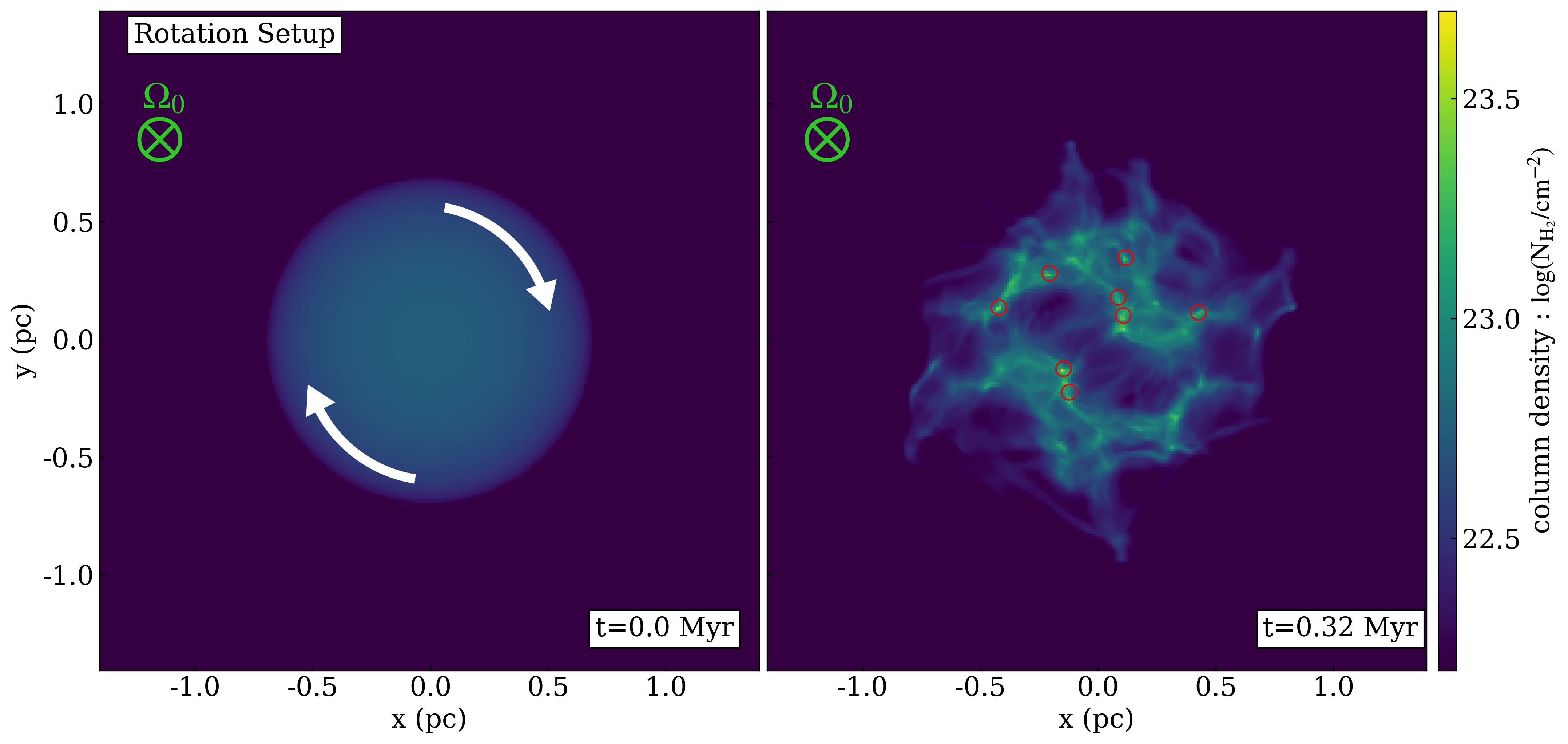}
  \includegraphics[width=13cm]{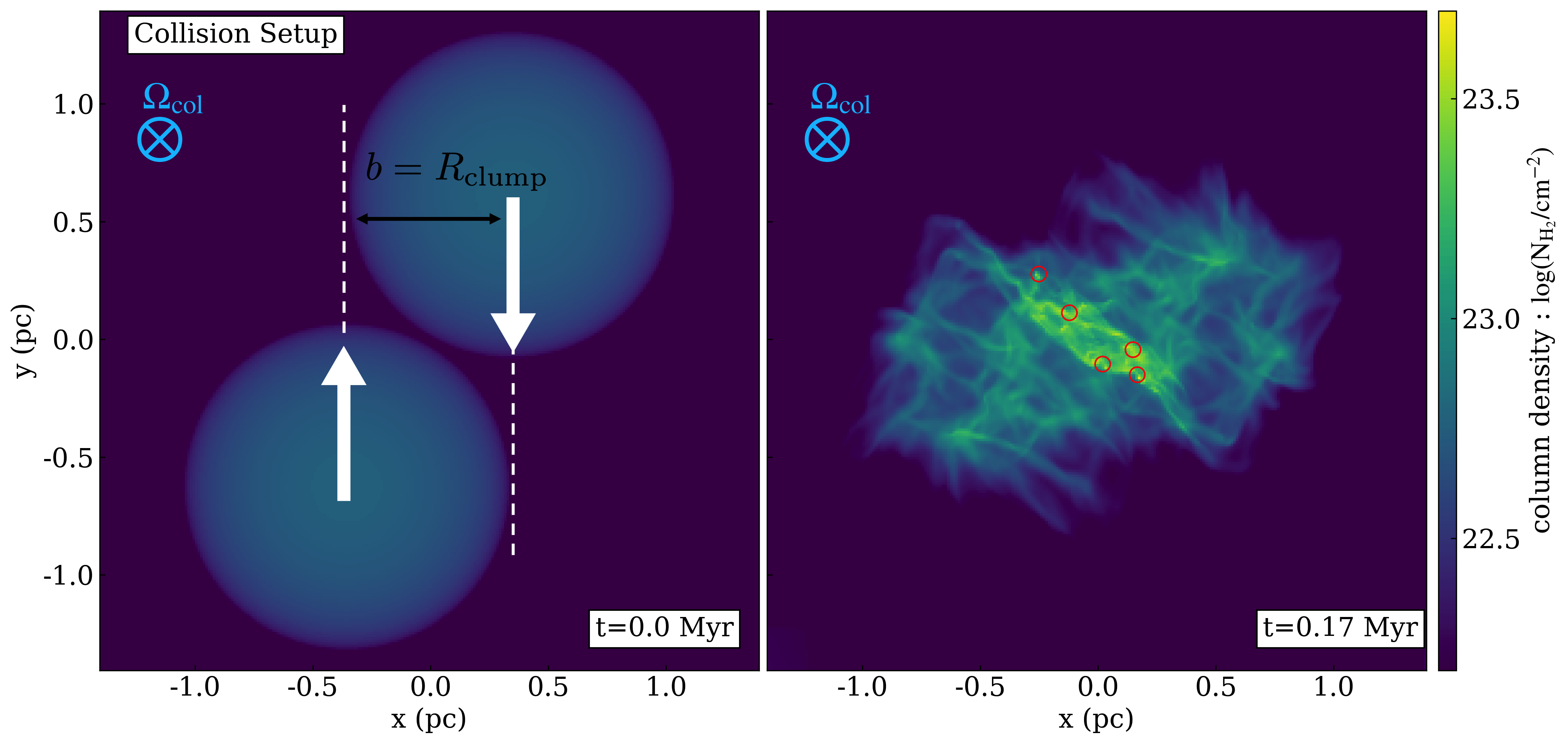}
\end{center}
\caption{Sample map from simulations considered in this study. The column density, as viewed along the $z$ axis, is shown. The top two rows display one of the Rotation Setup models ({\tt Rot-M5-B10P}) at $0.0$ and $0.32$ Myr. The orientation of the rotation axis of the clump is indicated by the symbol "$\otimes$" pointing in the direction of the $z$ axis. 
The bottom two rows display one of the Collision Setup models ({\tt Col-M5-B10P}) at $0.0$ and $0.17$ Myr. In the Collision Setup, the initial clump does not have an overall rotational velocity. However, after the collision, the two clumps rotate around their center due to the off-center configuration.
The positions of bound cores identified in simulations are additionally plotted as red open circles.}
\label{fig:initial_cindition}
\end{figure*}

The lower row of Figure \ref{fig:initial_cindition} shows sample maps from one of the Collision Setup model ({\tt Col-M5-B10P}). At the interface of the colliding, a high-density compressed layer is formed. Dense cores primarily form inside this shock layer. Two clump gas rotates around $\bm{\Omega}_{\rm col}$. In Appendix \ref{app:Time evolution of column density}, the time evolution of column density, as viewed along the $x$ axis, is shown.

\subsection{Measuring Core Properties}
\label{method:Measuring Core Properties}
For each set of model parameters, we typically conduct four simulations with different realizations of the input turbulence. The evolution of the clump was tracked to approximately the free-fall time $t_{\rm ff}\simeq 0.3 ~\rm Myr$. We identify gravitationally bound cores at the time when the most evolved core collapses ($n_{\rm max} \geqslant 10^{8}\rm ~cm^{-3}$) by applying the following criteria: (i) 
{$n\geq n_{\rm th}=10^{6}\rm ~cm^{-3}$}, (ii) cell number $>3^{3}$, (iii) total mass $M_{\rm core} >0.1M_{\odot}$, (iv) $E_{\rm thermal}+E_{\rm mag}+E_{\rm grav}<0 $ (details below). 
We tested multiple values of threshold density $n_{\rm th}$ and verified that the following results do not strongly depend on $n_{\rm th}$. We chose $n_{\rm th}=10^{6}\rm ~cm^{-3}$ since this is high enough to guarantee dense core formation but below $n_{\rm crit}$.

For some models, the total number of identified cores in the four simulations is below 20. In such cases, we perform additional simulations to ensure that the total number of cores exceeds 20, allowing for statistically meaningful discussions. In Section \ref{sec:Dynamics}, we also discuss unbound cores that do not satisfy the criterion (iv).

For each core, we calculated its thermal energy $E_{\rm thermal}$, kinetic energy $E_{\rm kin}$, magnetic energy, $E_{\rm mag}$, and self-gravitational energy $E_{\rm grav}$. $E_{\rm thermal}$ is given by 
\begin{equation}
E_{\rm thermal}=\sum_{i}\frac{3}{2}n_{i}kT_{i},
\end{equation}
where $i$ is an index of a cell in the core, $n_{i}$ is the number density, $k$ is the Boltzmann constant, and $T_{i}$ is the temperature.

$E_{\rm kin}$ is given by
\begin{equation}
E_{\rm kin}=\sum_{i}\frac{1}{2}\rho_{i}|\bm{v}_{i}-\bm{v}_{\rm mean}|^{2},
\end{equation}
where $\rho_{i}$ is the mass density, $\bm{v}_{i}$ is the velocity, and $\bm{v}_{\rm mean}$ is the mean velocity of core
defined by

\begin{equation}
\bm{v}_{\rm mean}=\frac{\sum_{i}\rho_{i}\bm{v}_{i}}{M_{\rm core}},
\end{equation}
where $M_{\rm core}$ is the mass of the core. $E_{\rm grav}$ is calculated by

\begin{equation}
E_{\rm grav}=-\frac{3GM_{\rm core}^2}{5R_{\rm core}},
\end{equation}
The core radius $R_{\rm core}$ is defined by
\begin{equation}
R_{\rm core}=\left(\frac{3V_{\rm core}}{4\pi}\right)^{1/3},
\end{equation}
where $V_{\rm core}$ is the total volume of the core.

$E_{\rm mag}$ is given by

\begin{equation}
E_{\rm mag}=\sum_{i}\frac{\bm{B}_{i}^2}{8\pi}\Delta V_{i},
\end{equation}
where $\bm{B}_{i}$ is the magnetic field flux density, and $\Delta V_{i}$ is the volume of a simulation cell. These calculation methods are similar to those used in earlier works \citep[e.g.,][]{2023MNRAS.522.4972S}.

We estimate the net angular momentum $\bm{L}_{\rm core}$ based on the calculation presented in \citet{Chen_2018}. $\bm{L}_{\rm core}$ is defined by the integration of each cell's relative angular momentum over the entire volume:
\begin{equation}
{\boldsymbol{L}}_{\rm core}=\sum_i \rho_i \Delta V_{i} \cdot\left(\boldsymbol{r}_i-\boldsymbol{r}_{\rm CM}\right) \times \boldsymbol{v}_i,
\end{equation}
where $\boldsymbol{r}_{\rm CM}$ is the center of mass. The rotational axis for the core is determined to be $\hat{\boldsymbol{L}}_{\rm core}=\bm{L}_{\rm core}/L_{\rm core}$,  where $L_{\rm core}=|{\boldsymbol{L}}_{\rm core}|$ is the magnitude of the net angular momentum of the core. 
The total rotational inertia of the core around this axis can be calculated by first determining the projected radius for each cell:
\begin{equation}
\boldsymbol{r}_{i, \perp}=\left(\boldsymbol{r}_i-\boldsymbol{r}_{\mathrm{CM}}\right)-\left[\left(\boldsymbol{r}_i-\boldsymbol{r}_{\mathrm{CM}}\right) \cdot \hat{\boldsymbol{L}}_{\rm core}\right] \hat{\boldsymbol{L}}_{\rm core}
\end{equation}

and then integrating over the whole volume:

\begin{equation}
I \equiv \sum_i \rho_i \Delta V_{i} \cdot\left|\boldsymbol{r}_{i, \perp}\right|^2
\end{equation}

The mean angular velocity $\Omega_{\rm core}$ and the rotational energy $E_{\rm rot}$ of the core are
\begin{gather}
\Omega_{\rm core} \equiv L_{\rm core} / I, \\
E_{\mathrm{rot}} \equiv \frac{1}{2} I \Omega_{\rm core}^2 .
\end{gather}

The mean magnetic field within the core is calculated by 

\begin{gather}
{\boldsymbol{B}}_{\rm core}=\frac{\sum_i \bm{B}_{i} \Delta V_{i} }{V_{\rm core}}. 
\end{gather}

In the following section, we will discuss the degree of alignment of $\bm{L}_{\rm core}$ or $\bm{B}_{\rm core}$. 
To quantify the alignment level and its significance, we use the orientation parameter $S$: 

\begin{equation}
\label{eq: orientation parameter}
S=\frac{3\left\langle\cos ^2 \chi\right\rangle-1}{2},
\end{equation}
where $\chi$ is the angle with respect to the director. 
In the case of a perfect alignment, $S=1$, while in the case of a completely random alignment, $S=0$. 
When $0 < S < 1$, it denotes a partial alignment. For example, when $\left\langle\cos ^2 \chi\right\rangle=\cos ^2 45^{\circ}$, $S=0.25$. In the following discussion, $S>0.25$ will be referred to as strong alignment and $0.25>S>0$ as weak alignment.

\begin{deluxetable}{lccccc}
\tablecaption{Summary of Simulations and Explored Parameter Space}
\label{tab:parameter}
\tablewidth{0pt}
\tablehead{
\colhead{Model name \hspace{40pt} } & \colhead{$V_{0}\,^{\rm a}$} & \colhead{$\mathcal{M}\,^{\rm b}$} & \colhead{$B_{0}\,^{\rm c}$}&
\colhead{$\theta_{0}\,^{\rm d}$} & \colhead{$\alpha\,^{\rm e}$} \\
\nocolhead{} & \colhead{($\rm km~s^{-1}$)} & \nocolhead{} & \colhead{($\mu G$)} &
\colhead{($^{\circ}$)} & \nocolhead{}
}
\startdata
Rotation Setup              &      &          &         &              & \\ \hline
\tt{Rot-M1.5-B10P}          & \nodata  & 1.5      & 10      & 0        &  0.30   \\
\tt{Rot-M1.5-B100P}         & \nodata  & 1.5      & 100     & 0        &  0.53   \\
\tt{Rot-M5-B10P}            & \nodata  & 5        & 10      & 0        &  0.53   \\
\tt{Rot-M5-B100P}           & \nodata  & 5        & 100     & 0        &  0.75   \\
\tt{Rot-M1.5-B10D}          & \nodata  & 1.5      & 10      & 45       &  0.30       \\
\tt{Rot-M1.5-B100D}         & \nodata  & 1.5      & 100     & 45       &  0.53       \\
\tt{Rot-M5-B10D}            & \nodata  & 5        & 10      & 45       &  0.53       \\
\tt{Rot-M5-B100D}           & \nodata  & 5        & 100     & 45       &  0.75       \\ \hline
w/o Setup$\,^{\rm f}$                   &      &          &         &    &      \\ \hline
\tt{w/o-M1.5-B10}           &  \nodata & 1.5      & 10      &  \nodata &  0.05   \\
\tt{w/o-M1.5-B100}          &  \nodata & 1.5      & 100     &  \nodata &  0.28   \\
\tt{w/o-M5-B10}             &  \nodata & 5        & 10      &  \nodata &  0.28   \\
\tt{w/o-M5-B100}            &  \nodata & 5        & 100     &  \nodata &  0.50   \\\hline
Collision Setup             &      &          &         &              &         \\ \hline
\tt{Col-M1.5-B10P}          & 2.8  & 1.5      & 10      & 0            &  0.05  \\
\tt{Col-M1.5-B100P}         & 2.8  & 1.5      & 100     & 0            &  0.28  \\
\tt{Col-M5-B10P}            & 2.8  & 5        & 10      & 0            &  0.28  \\
\tt{Col-M5-B100P}           & 2.8  & 5        & 100     & 0            &  0.50   \\
\tt{Col-M1.5-B10D}          & 2.8  & 1.5      & 10      & 45           &  0.05   \\
\tt{Col-M1.5-B100D}         & 2.8  & 1.5      & 100     & 45           &  0.28   \\
\tt{Col-M5-B10D}            & 2.8  & 5        & 10      & 45           &  0.28   \\
\tt{Col-M5-B100D}           & 2.8  & 5        & 100     & 45           &  0.50    \\
\tt{Col-S-M1.5-B10P}        & 1.4  & 1.5      & 10      & 0            &  0.05   \\
\tt{Col-S-M1.5-B100P}       & 1.4  & 1.5      & 100     & 0            &  0.28   \\
\tt{Col-S-M5-B10P}          & 1.4  & 5        & 10      & 0            &  0.28   \\
\tt{Col-S-M5-B100P}         & 1.4  & 5        & 100     & 0            &  0.50    \\ \hline
\enddata
\tablecomments{$^{\rm a}$ The pre-collision velocity of the clump. 
$^{\rm b}$ The Mach number of turbulence.
$^{\rm c}$ The strength of initial magnetic field.
$^{\rm d}$ The angle between the initial magnetic field $\bm{B}_{0}$ relative to the $\bm{\Omega}_{0}$ ($\bm{\Omega}_{\rm col}$).
$^{\rm e}$ Energy ratio of sum of turbulent, magnetic field, rotation, and thermal energies to absolute value of self-gravitational energy for the initial initial clump, $(E_{\rm clump,tur}+E_{\rm clump,mag}+E_{\rm clump,rot}+E_{\rm clump,therm})/|E_{\rm clump,grav}|$.
$^{\rm f}$ The models of the clump that has neither initial angular momentum with $\Omega_{0}$ nor collide. 
}
\end{deluxetable}

\section{Results}
\label{sec:Results}

Overall, we compare the results of 24 different simulation models listed in Table \ref{tab:parameter}. For each set of model parameters, we run multiple simulations with different realizations of the input turbulence, and identify bound cores. We present analysis result of bound cores in the Rotation Setup in Section \ref{sec:result_Rotation Setup}, and those of the Collision Setup in Section \ref{sec:result_collision Setup}. In Appendix \ref{app:property}, Table \ref{tab:Properties of identified cores} summarizes the physical properties measured from cores. In Appendix \ref{app:Nearest neighbor core separation}, we show distributions of nearest-neighbor core separations.

\subsection{Rotation Setup}
\label{sec:result_Rotation Setup}
This section shows the simulation results of the Rotation Setup. 
For each identified core, we measured the net angular momentum $\bm{L}_{\rm core}$ and mean magnetic field $\bm{B}_{\rm core}$. We show the correlation between $\bm{L}_{\rm core}$ and the rotational axis of the clump, $\bm{\Omega}_{0}$, in Section \ref{sec:rot_Angular momentum}. In Section \ref{sec:rot_Magnetic field}, we discuss the alignment of $\bm{B}_{\rm core}$. 
In Section \ref{sec:rot_Rotation-Magnetic field relation}, we present the rotation-magnetic field relation among bound cores.

\subsubsection{Angular Momentum of the Rotation Setup}
\label{sec:rot_Angular momentum}

\begin{figure}
\begin{center}
  \includegraphics[width=8.5cm]{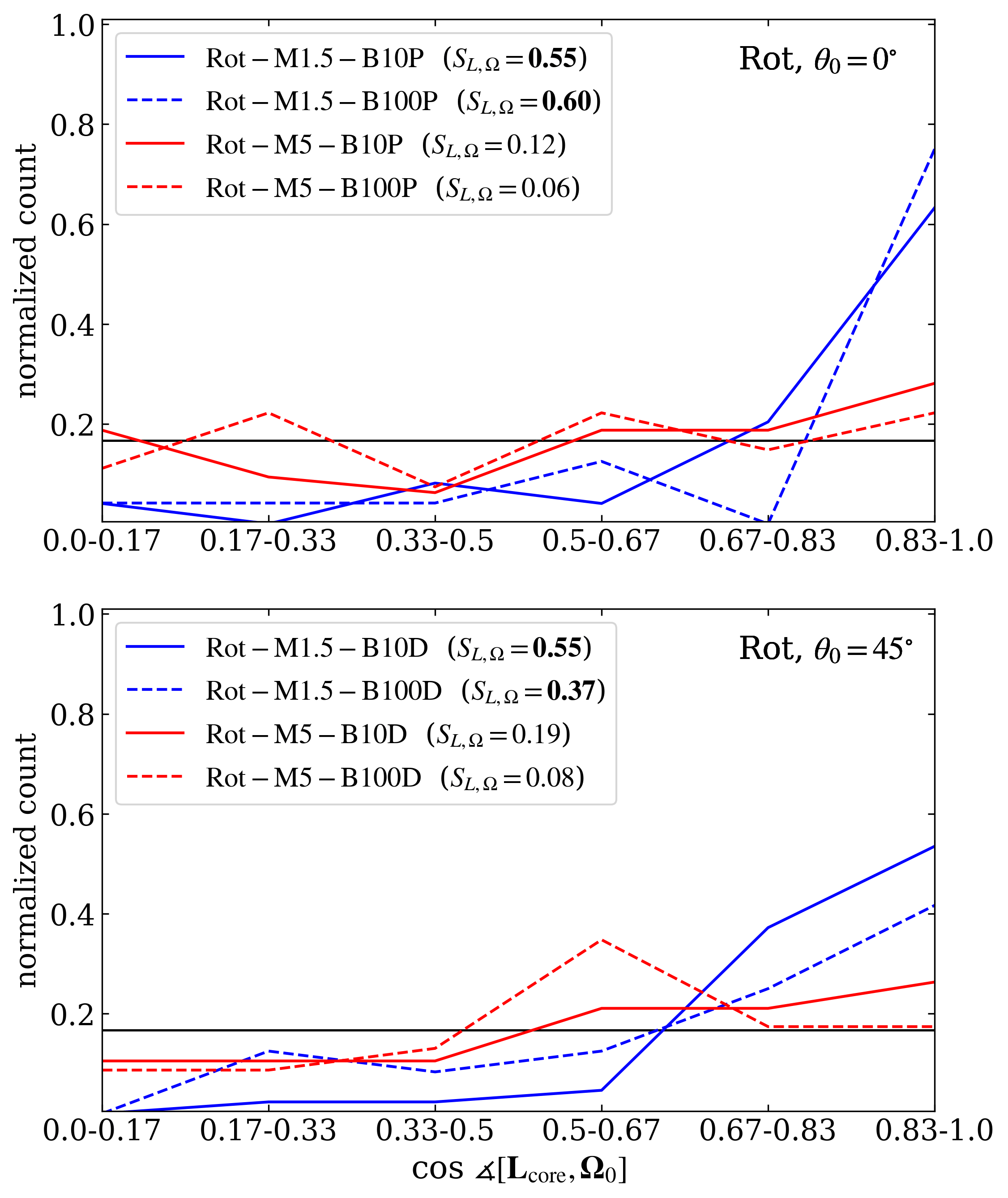}
\end{center}
\caption{Histograms of the cosine of the relative angles between parental clump rotation axis $\bm{\Omega}_{0}$ and the integrated angular momentum $\bm{L}_{\rm core}$, for all bound cores formed in different models. The black line shows the expected distribution for an isotropic orientation of $\bm{L}_{\rm core}$. In the legends, the orientation parameter $S_{L,\Omega}=(3\langle \mathrm{cos}^{2} \measuredangle[\bm{L}_{\rm core},\bm{\Omega}_{0}]\rangle-1)/2$ for each model are shown, and values with $S_{L,\Omega}>0.25$ are indicated in bold. 
$\theta_{0}=0^{\circ}$ cases are shown in the first row, while $\theta_{0}=45^{\circ}$ cases are shown in the second row. In all models with weak turbulence (indicated by blue lines), there is a clear tendency for $\bm{L}_{\rm core}$ and $\bm{\Omega}_{0}$ to align. On the other hand, in the strong turbulence models (indicated by red lines), the angles are close to being isotropically distributed and no tendency for alignment is observed.}
\label{fig:rot_L_omega}
\end{figure}

\begin{figure}
\begin{center}
  \includegraphics[width=7.0cm]{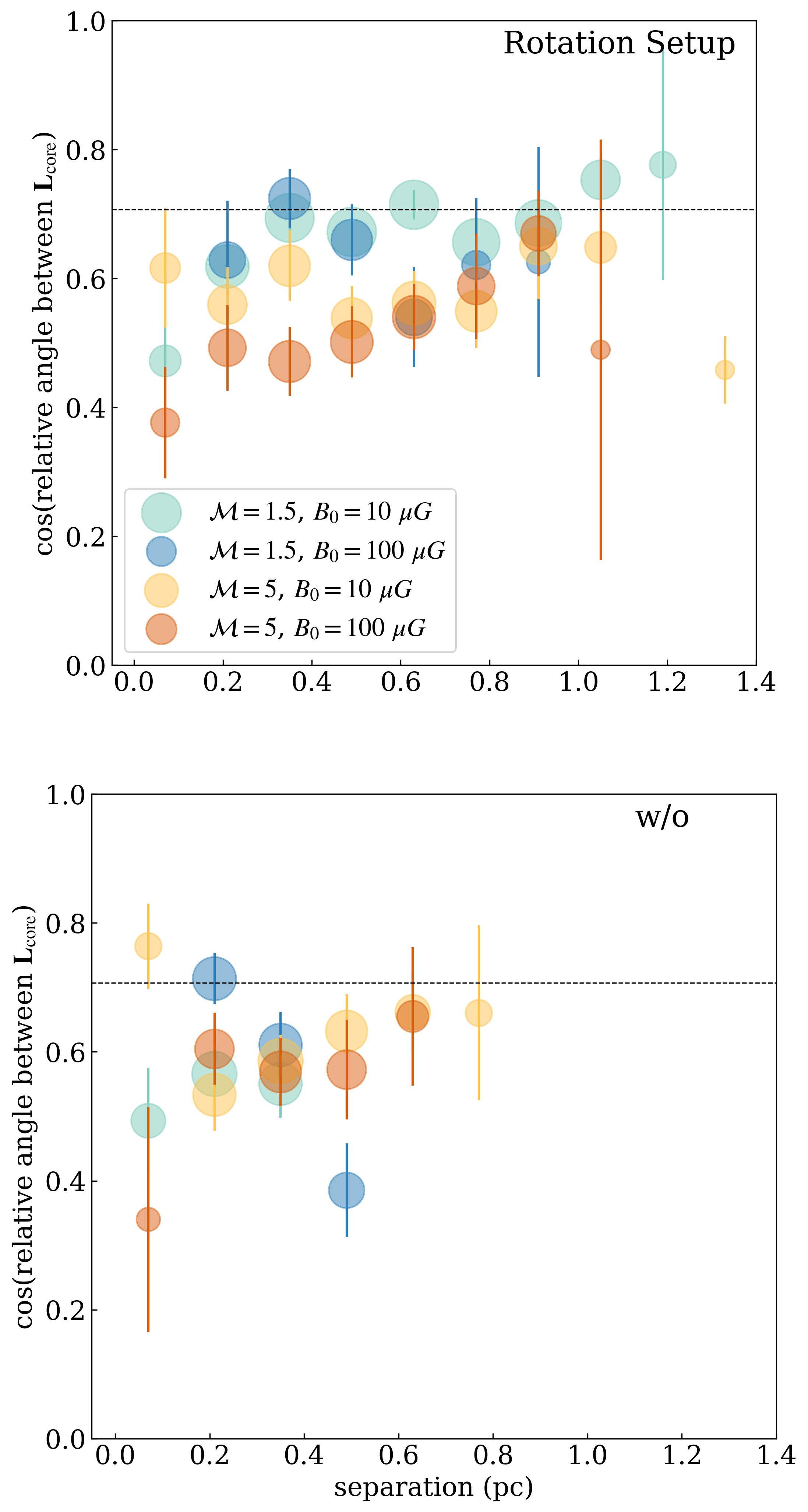}
\end{center}
\caption{The cosine of relative orientation angles of $\bm{L}_{\rm core}$ pairs as a function of their separation distances. The data are averaged in various separation distance bins, with the size of the circle representing the number of samples in each distance bin. The black dashed line indicates $\mathrm{cos\,45^{\circ}}$. The models with $\theta_{0}=0^{\circ}$ and $45^{\circ}$ are presented together. Rotation Setup is shown in the top panel and w/o Setup in the bottom panel. In the Rotation Setup, a stronger alignment of $\bm{L}_{\rm core}$ pairs is observed, especially when the turbulence is weak ($\mathcal{M}=1.5$). }
\label{fig:rot_L_pair}
\end{figure}

We examine the orientation of $\bm{L}_{\rm core}$ for all bound cores in different models to determine if the cores reflect the average angular momentum of the clump.
Figure \ref{fig:rot_L_omega} illustrates the histograms of the cosine of the angle between parental clump rotation axis $\bm{\Omega}_{0}$ and the core angular momentum $\bm{L}_{\rm core}$. Here, the orientation parameter $S_{L,\Omega}$ is defined as $S_{L,\Omega}=(3\langle \mathrm{cos}^{2} \measuredangle[\bm{L}_{\rm core},\bm{\Omega}_{0}]\rangle-1)/2$. In all models with weak turbulence (indicated by blue lines), strong alignment ($S_{L,\Omega}>0.25$) is achieved regardless of the strength or orientation of the initial magnetic field $\bm{B}_{0}$. For these models, the distribution of $\mathrm{cos}\,\measuredangle[\bm{L}_{\rm core},\bm{\Omega}_{0}]$ deviates greatly from an uniform distribution and the null hypothesis that ``the distribution is uniform'' is rejected at a significance level of 5\% using the Kolmogorov-Smirnov (K-S) test. In models with weak turbulence, the rotation of the parental clump is passed down to the bound cores,
while in models with strong turbulence (indicated by red lines), the distribution of spin axes is found to be close to isotropic, and the null hypothesis that "the distribution is uniform" cannot be rejected. This suggests that the rotational motion of the clumps is not reflected in the bound cores in strong turbulence models. Therefore, the turbulence intensity is a critical parameter that determines whether the rotational motion of the parental clump is inherited by the bound cores or not. Expressed using the ratio of $E_{\rm clump, rot}$ to $E_{\rm clump, tur}$ of the clump, there is no tendency for alignment when $E_{\rm clump, rot}$/$E_{\rm clump,tur}\sim 1$, and a strong tendency for alignment when $E_{\rm clump, rot}$/$E_{\rm clump,tur}>1$.  

Another important point to note is that in the case of weak turbulence, $\bm{L}_{\rm core}$ tends to align with $\bm{\Omega}_{0}$ regardless of the strength $B_{0}$ or orientation $\theta_{0}$ of $\bm{B}_{0}$. As shown later in Section \ref{sec:rot_Rotation-Magnetic field relation}, there is almost no correlation between $\bm{L}_{\rm core}$ and $\bm{B}_{\rm core}$. At least within the range of parameters investigated in this study, the magnetic field does not constrain the direction of $\bm{L}_{\rm core}$.

Between all $\bm{L}_{\rm core}$ pairs in each simulation run, we calculate the cosine of the relative orientation angle, noted as $\mathrm{cos}\,\theta_{L,L}$ and their separations. Figure \ref{fig:rot_L_pair} displays $\mathrm{cos}\,\theta_{L,L}$ as a function of their separation distances between all pairs of $\bm{L}_{\rm core}$. The results are binned by separation distances of pairs and presented as the mean and standard deviation for each bin. The top panel shows the results of the Rotation Setup, while the bottom panel shows those of w/o Setup for comparison. In the Rotation Setup, particularly with weak turbulence ($\mathcal{M}=1.5$), $\mathrm{cos}\,\theta_{L,L}$ is generally close to $\mathrm{cos}\,45^{\circ}$ indicating a relatively stronger alignment of $\bm{L}_{\rm core}$ pairs. Weak turbulence models exhibit relatively high $\mathrm{cos}\,\theta_{L,L}$ values over a wide range of separations from 0.2 to 1.2 pc, suggesting that $\bm{L}_{\rm core}$ pairs are generally aligned regardless of the distance between cores. As shown in Figure \ref{fig:rot_L_omega}, in the Rotation Setup with the weak turbulence, $\bm{L}_{\rm core}$ inherits the rotation of clumps. Therefore, $\bm{L}_{\rm core}$ tends to align with similar angles between pairs. On the other hand, in Rotation Setup with the strong turbulence ($\mathcal{M}=5$), $\mathrm{cos}\,\theta_{L,L}$ is close to $\sim$ 0.5 which would be expected from a uniform distribution. As shown in Figure \ref{fig:rot_L_omega}, in the case of strong turbulence, the rotation of the clump is not transferred to the core, so the pairs of $\bm{L}_{\rm core}$ did not align. In the case of w/o Setup, since the clump is not rotating globally, the direction of each $\bm{L}_{\rm core}$ is random, and $\mathrm{cos}\,\theta_{L,L}$ is around 0.5 at any separation, indicating no alignment between $\bm{L}_{\rm core}$ pairs.

\subsubsection{Magnetic Field of the Rotation Setup}
\label{sec:rot_Magnetic field}

\begin{figure}
\begin{center}
  \includegraphics[width=8.5cm]{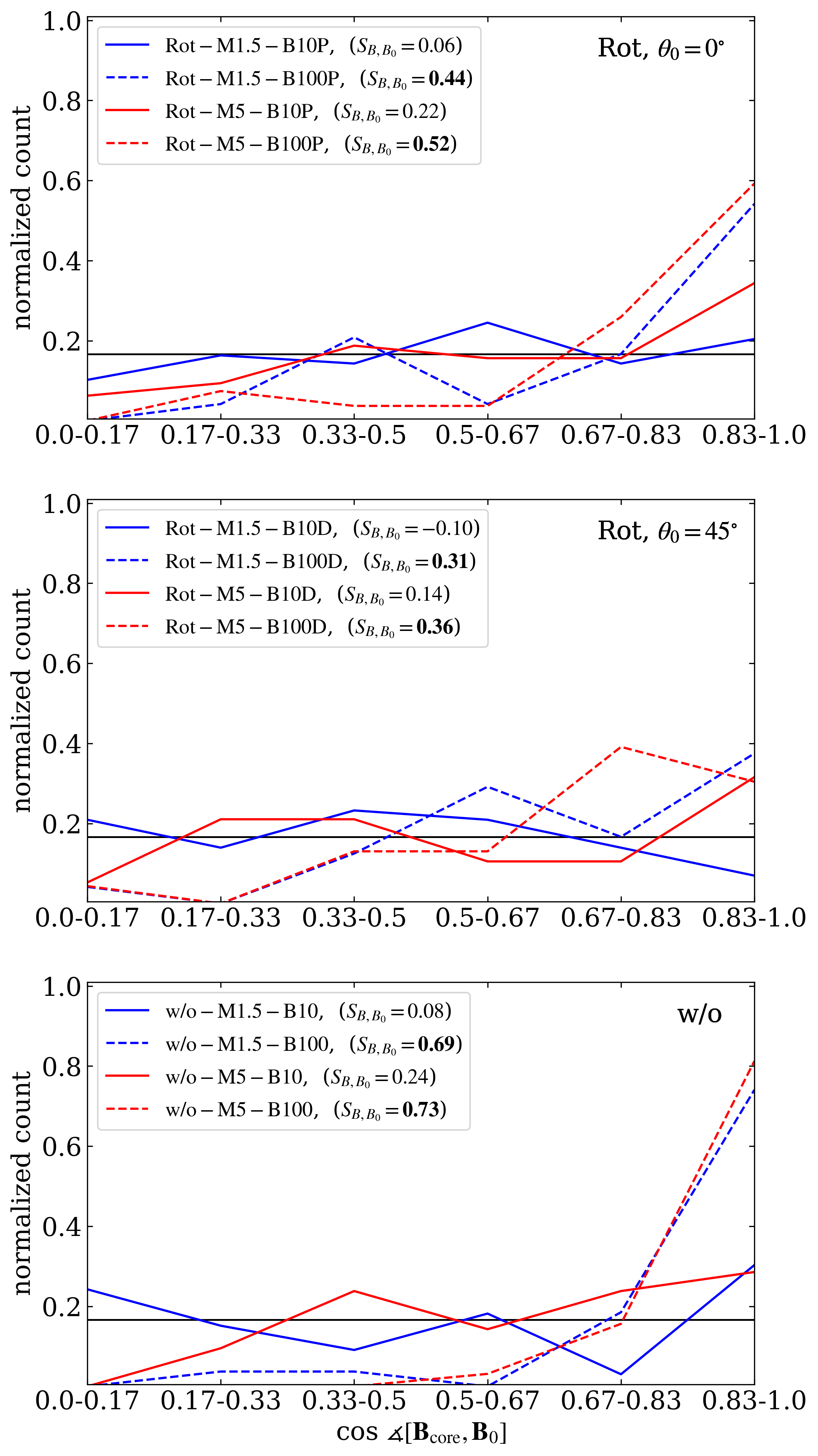}
\end{center}
\caption{Histograms of the cosine of the relative angle between the initial magnetic field $\bm{B}_{0}$ and the mean magnetic field within the core, $\bm{B}_{\rm core}$, for bound cores formed in different models. The black line shows the expected distribution for an isotropic orientation of $\bm{B}_{\rm core}$. In the legends, the orientation parameter $S_{B,B_{0}}=(3\langle \mathrm{cos}^{2} \measuredangle[\bm{B}_{\rm core},\bm{B}_{0}]\rangle-1)/2$ for each model are shown, and values with $S_{L,B_{0}}>0.25$ are indicated in bold. 
Rotation setup models with $\theta_{0}=0^{\circ}$ are shown in the top panel, while $\theta_{0}=45^{\circ}$ models are shown in the middle panel. w/o Setup models are shown in the bottom panel. Generally, in models with strong magnetic fields (indicated by dashed lines), the strong alignment of $\bm{B}_{\rm core}$ with $\bm{B}_{0}$ is observed.}
\label{fig:rot_B_B}
\end{figure}

\begin{figure}
\begin{center}
  \includegraphics[width=7.0cm]{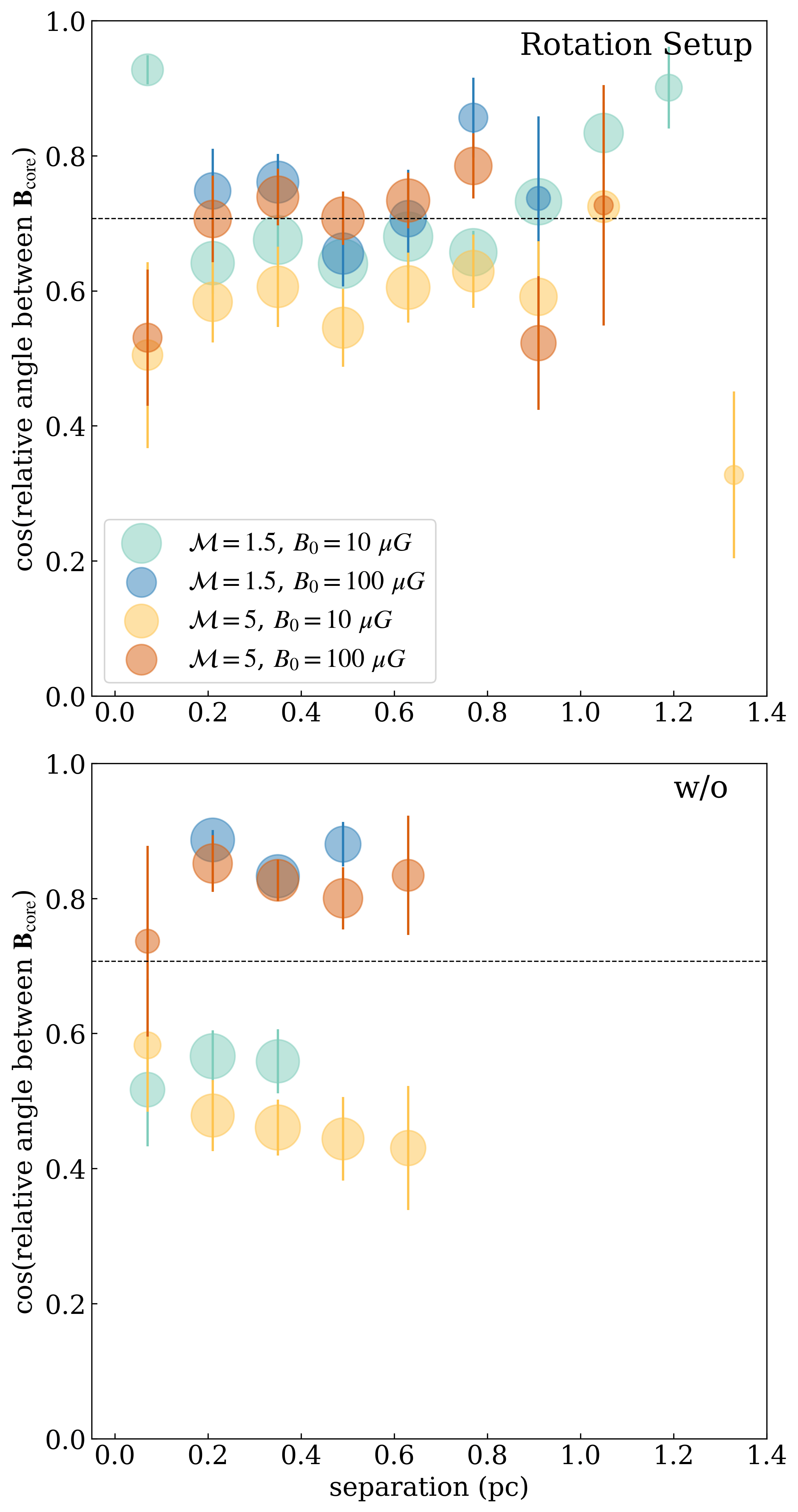}
\end{center}
\caption{The cosine of relative orientation angles of $\bm{B}_{\rm core}$ pairs as a function of their separation distances. The presentation method is similar to Figure \ref{fig:rot_L_pair}.
Models with strong magnetic fields ($B_{0}=100\,\mu G$) have a higher degree of alignment of $\bm{B}_{\rm core}$ pairs compared to models with weak magnetic fields ($B_{0}=10\,\mu G$).}
\label{fig:rot_B_pair}
\end{figure}

We explored the orientation of $\bm{B}_{\rm core}$ for all bound cores and investigated the correlation between  $\bm{B}_{\rm core}$ and the initial magnetic field of the clump, $\bm{B}_{0}$. 
Figure \ref{fig:rot_B_B} shows the histograms of cosine of the angle between $\bm{B}_{\rm core}$ and $\bm{B}_{0}$. The parameter $S_{B,B_{0}}$ is defined as $(3\langle \mathrm{cos}^{2} \measuredangle[\bm{B}_{\rm core},\bm{B}_{0}]\rangle-1)/2$. In all models with strong ${B}_{0}$ (indicated by dashed lines), $S_{B,B_{0}}$ is larger than 0.25 indicating the strong alignment between  $\bm{B}_{\rm core}$ and $\bm{B}_{0}$. The null hypothesis that ``the distribution is uniform'' is rejected at a significance level of 5\% for all models with strong ${B}_{0}$. Strong magnetic fields tend to maintain their coherence along the initial direction. Therefore, $\bm{B}_{\rm core}$ inherits the initial orientation of the clump's field, which leads to a tendency for $\bm{B}_{\rm core}$ to align parallel to $\bm{B}_{0}$. Also, for some models with weak ${B}_{0}$, $S_{B,B_{0}}$ is positive and the null hypothesis that ``the distribution is uniform'' is rejected. However, the degree of alignment is weaker compared to models of strong ${B}_{0}$. This is due to the fact that when the magnetic field is weak, the magnetic field inside the clump is easily disturbed by turbulence or rotational motion of the clump, resulting in the misalignment of $\bm{B}_{\rm core}$. 

Similar tendencies are obtained through the analysis of $\bm{B}_{\rm core}$ pairs. Figure \ref{fig:rot_B_pair} shows the cosine of the relative orientation angle, noted
as $\mathrm{cos}\,\theta_{B,B}$, between all $\bm{B}_{\rm core}$ pairs as a function of their separation distances in the same manner as Figure \ref{fig:rot_L_pair}. 
In the models of Rotation Setup with a strong $B_{0}$ ($=100\,\mu G$), there is a higher degree of alignment with an average $\mathrm{cos}\,\theta_{B,B}$ around $\mathrm{cos}\,45^{\circ}$ over a wide range of separations from 0.2 to 1.2 pc. 
On the other hand, in the weak $B_{0}$ ($=10\,\mu G$) models of Rotation Setup, the degree of alignment is generally lower compared to the strong ${B}_{0}$ models for a range of separations between 0.2 and 0.8. The strength of the initial magnetic field determines the degree of alignment of $\bm{B}_{\rm core}$ pairs.

In w/o Setup, the difference in results between strong and weak ${B}_{0}$ cases is significant. In the weak ${B}_{0}$ models, $\cos\,\theta_{B,B}$ is around 0.5, while in the strong ${B}_{0}$ cases, the degree of alignment of $\bm{B}_{\rm core}$ pairs is higher and surpasses $\mathrm{cos}\,45^{\circ}$.

\subsubsection{Rotation-Magnetic field relation of the Rotation Setup}
\label{sec:rot_Rotation-Magnetic field relation}

\begin{figure}
\begin{center}
  \includegraphics[width=5.85cm]{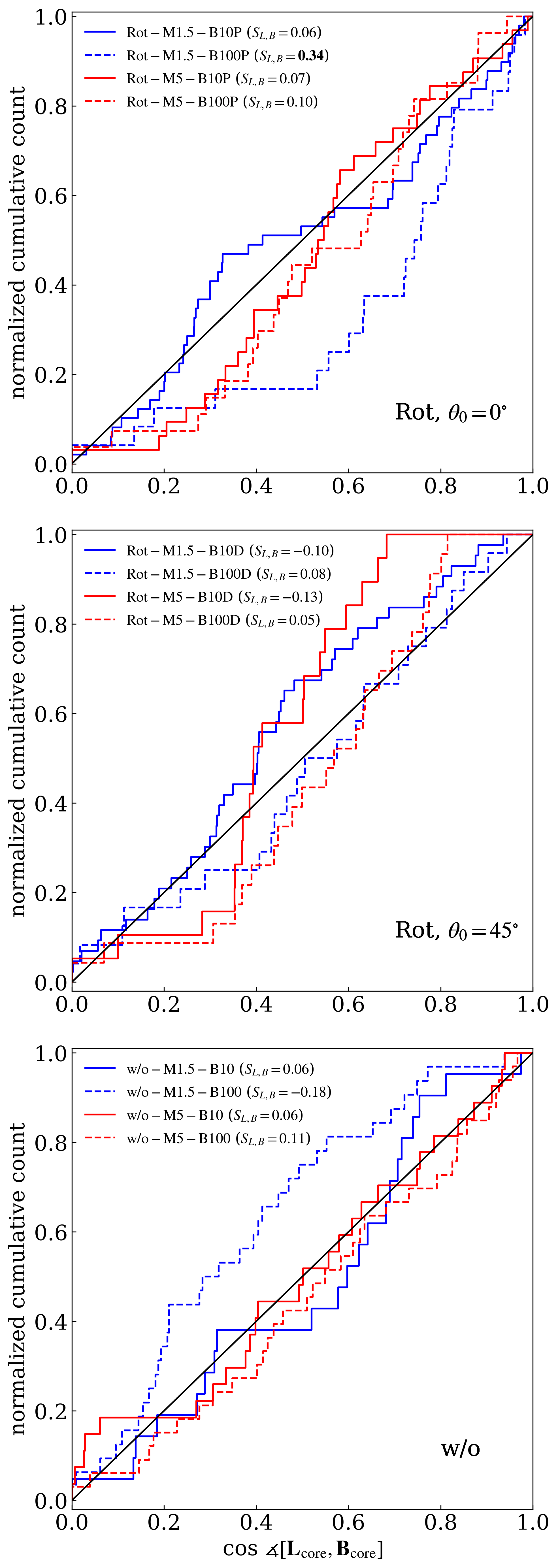}
\end{center}
\caption{The cumulative distribution function of the cosine of the relative angle between the mean magnetic field $\bm{B}_{\rm core}$ and the angular momentum $\bm{L}_{\rm core}$, compared to the expected CDF (black lines) for a completely uniform distribution. Rotation Setup models with $\theta_{0}=0^{\circ}$ are shown in the top panel, while $\theta_{0}=45^{\circ}$ models are shown in the middle panel. The bottom panel shows the results of w/o Setup. In most models, suggesting random distributions of $\measuredangle[\bm{L}_{\rm core},\bm{B}_{\rm core}]$.}
\label{fig:rot_LB}
\end{figure}

We investigated the relation between $\bm{L}_{\rm core}$ and $\bm{B}_{\rm core}$. Figure \ref{fig:rot_LB} illustrates the cumulative distribution function (CDF) of the cosine of the relative angle between $\bm{L}_{\rm core}$ and $\bm{B}_{\rm core}$ compared to a uniform distribution. The orientation parameter $S_{L,B}$ is defined as $S_{L,B}=(3\langle \mathrm{cos}^{2} \measuredangle[\bm{L}_{\rm core},\bm{B}_{\rm core}]\rangle-1)/2$. The CDF of most models appears to be a relatively straight line and the null hypothesis that ``the distribution is uniform'' is not rejected at a significance level of 5 \% using the K-S test except for the model {\tt Rot-M1.5-B100P}. 

The model {\tt Rot-M1.5-B100P} has weak initial turbulence intensity ($\mathcal{M}=1.5$), resulting in a well-aligned $\bm{L}_{\rm core}$ with $\bm{\Omega}_{0}$, as shown in Section \ref{sec:rot_Angular momentum}. Furthermore, the strong initial magnetic field ($B_{0}=100\mu G$) contributes to the alignment between $\bm{B}_{\rm core}$ and $\bm{B}_{0}$ (see Section \ref{sec:rot_Magnetic field}), and as $\theta_{0}=0^{\circ}$ ($\bm{\Omega}_{0} \parallel \bm{B}_{0}$), $\bm{L}_{\rm core}$ is well aligned with $\bm{B}_{\rm core}$. As such, due to the limited initial conditions and geometric reasons, $\bm{L}_{\rm core}$ and $\bm{B}_{\rm core}$ are aligned, resulting in a high orientation parameter $S_{L,B}=0.34$. In models other than {\tt Rot-M1.5-B100P} with such specific initial conditions, the distribution of angle between $\bm{L}_{\rm core}$ and $\bm{B}_{\rm core}$ is generally random.
We can  conclude that $\bm{B}_{\rm core}$ does not strongly limit $\bm{L}_{\rm core}$. As shown in Section \ref{sec:rot_Angular momentum}, it is the rotation of clumps and turbulence that determine the property of $\bm{L}_{\rm core}$.

\subsection{Collision Setup}
\label{sec:result_collision Setup}
This section shows the simulation results of the Collision Setup. 
We show the correlation between the $\bm{L}_{\rm core}$ and $\bm{\Omega}_{\rm col}$ in Section \ref{sec:col_Angular momentum}. In Section \ref{sec:col_Magnetic field}, we discuss the alignment of $\bm{B}_{\rm core}$. In Section \ref{sec:col_Rotation-Magnetic field relation}, we present the rotation-magnetic field relation among bound cores.

\subsubsection{Angular Momentum of the Collision Setup}
\label{sec:col_Angular momentum}

\begin{figure}
\begin{center}
  \includegraphics[width=8.5cm]{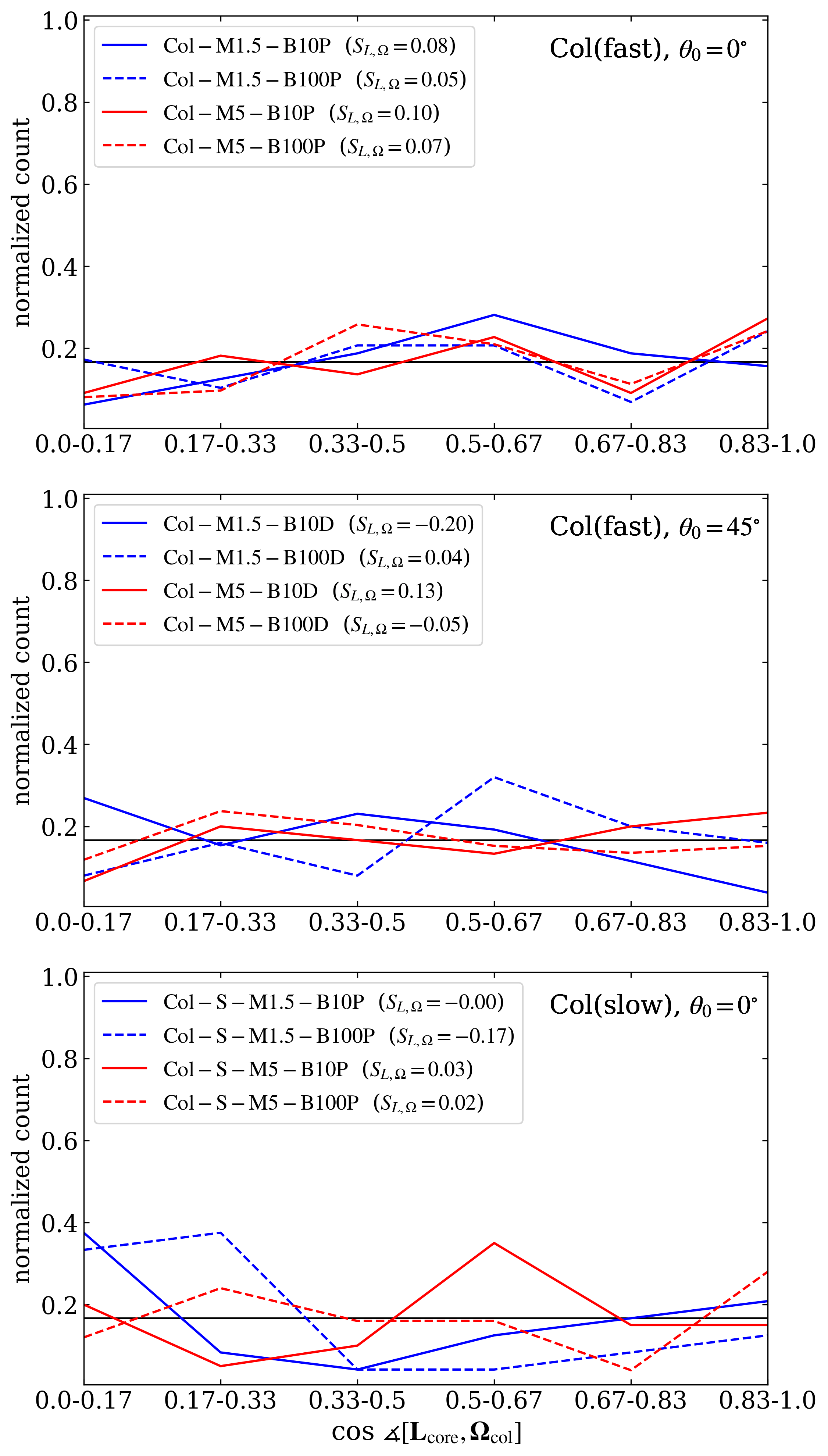}
\end{center}
\caption{Same as Figure \ref{fig:rot_L_omega} except for Collision Setup models. Fast collision velocity cases with $\theta_{0}=0^{\circ}$ are shown in the top panel and those of $\theta_{0}=45^{\circ}$ are shown in the middle panel. Slow collision velocity cases are shown in the bottom panel. For all models in the Collision Setup, the angles are close to being isotropically distributed and no tendency for alignment is observed.}
\label{fig:col_L_omega}
\end{figure}

\begin{figure}
\begin{center}
  \includegraphics[width=7.0cm]{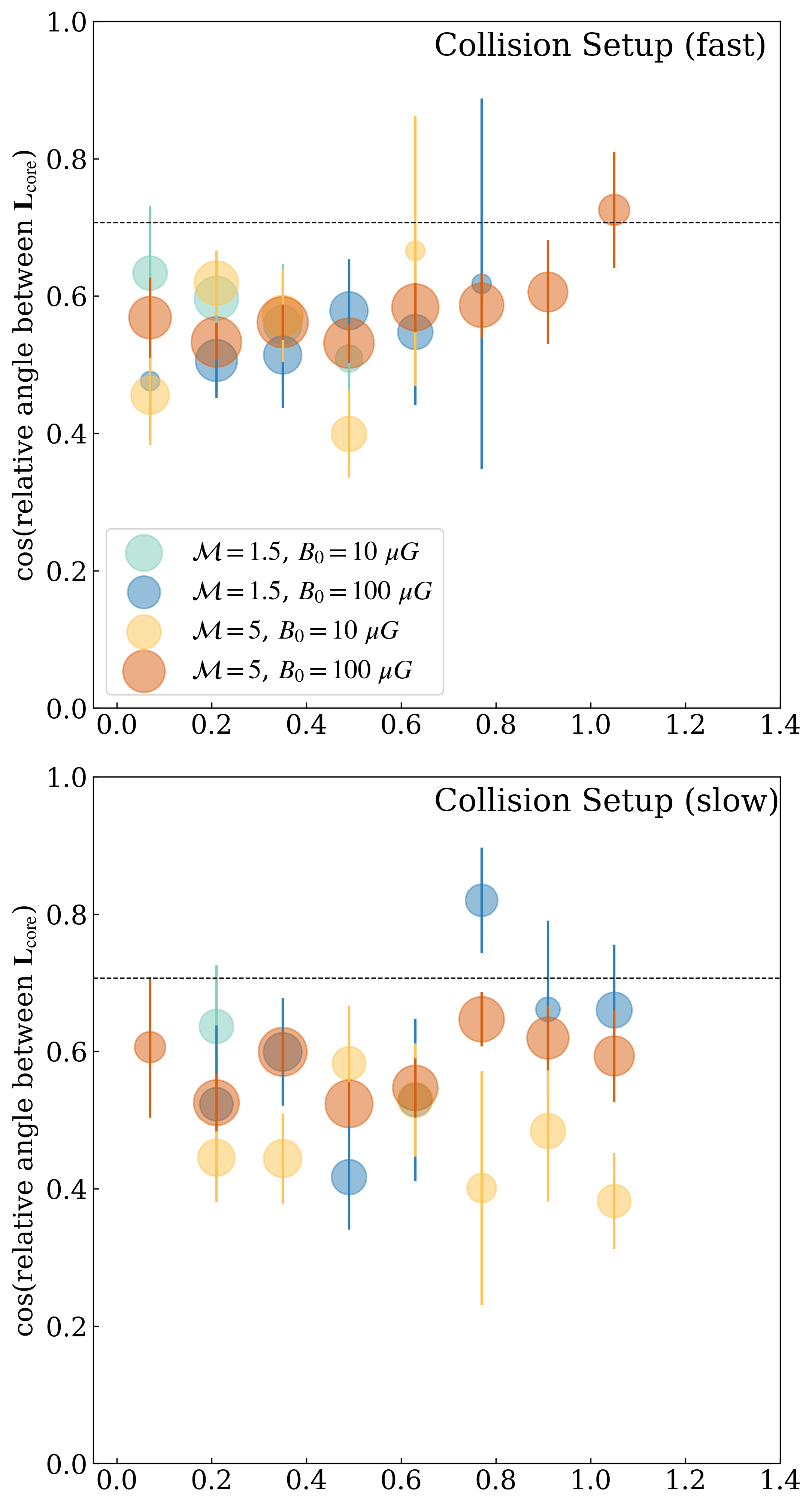}
\end{center}
\caption{Same as Figure \ref{fig:rot_L_pair} except for Collision Setup models. Fast collision velocity cases are shown in the top panel, and slow collision velocity cases are shown in the bottom panel. Regardless of the models, $\mathrm{cos},\theta_{L,L}$ is close to 0.5 for most separation ranges, which suggests that the direction of $\bm{L}_{\rm core}$ is randomly distributed.}
\label{fig:col_L_pair}
\end{figure}

In the Collision Setup, the initial clumps are set to have no initial rotational angular velocity ${\Omega}_{0}$. Nonetheless, due to their off-center arrangement, the two clumps begin to rotate after the collision, resulting in a dominant momentum in the plane perpendicular to the rotation axis $\bm{\Omega}_{\rm col}$. It is therefore expected that the gas motion of the clumps would be inherited by the cores formed, with the angular momentum vector $\bm{L}_{\rm core}$ aligning with $\bm{\Omega}_{\rm col}$ (see Figure \ref{fig:L_z_time}). However, contrary to this expectation, the analysis results reveal a different outcome.

Figure \ref{fig:col_L_omega} illustrates the histograms of cosine of the angle between $\bm{L}_{\rm core}$ and $\bm{\Omega}_{\rm col}$ as Figure \ref{fig:rot_L_omega}. Here, the orientation parameter $S_{L,\Omega}$ is defined as $S_{L,\Omega}=(3\langle \mathrm{cos}^{2} \measuredangle[\bm{L}_{\rm core},\bm{\Omega}_{\rm col}]\rangle-1)/2$. 
In the Collision Setup, $S_{L,\Omega}$ is generally small regardless of turbulence strength. Also, all Collision Setup models are roughly consistent with a completely uniform distribution of $\mathrm{cos}\,\measuredangle[\bm{L}_{\rm core},\bm{\Omega}_{\rm col}]$, which cannot be rejected at a significance level of 5\%. In the Rotation Setup, a clear alignment tendency was observed between $\bm{L}_{\rm core}$ and $\bm{\Omega}_{0}$ for weak turbulence. However, in the Collision Setup, the clump's rotation is not transferred to the core regardless of the initial parameters.

Figure \ref{fig:col_L_pair} displays $\mathrm{cos}\,\theta_{L,L}$ as a function of their separation distances as Figure \ref{fig:rot_L_pair}. For any models, $\mathrm{cos}\,\theta_{L,L}$ is around 0.5 for most separation ranges, indicating the random distribution of the direction of $\bm{L}_{\rm core}$.

\subsubsection{Magnetic Field of the Collision Setup}
\label{sec:col_Magnetic field}

\begin{figure}
\begin{center}
  \includegraphics[width=8.5cm]{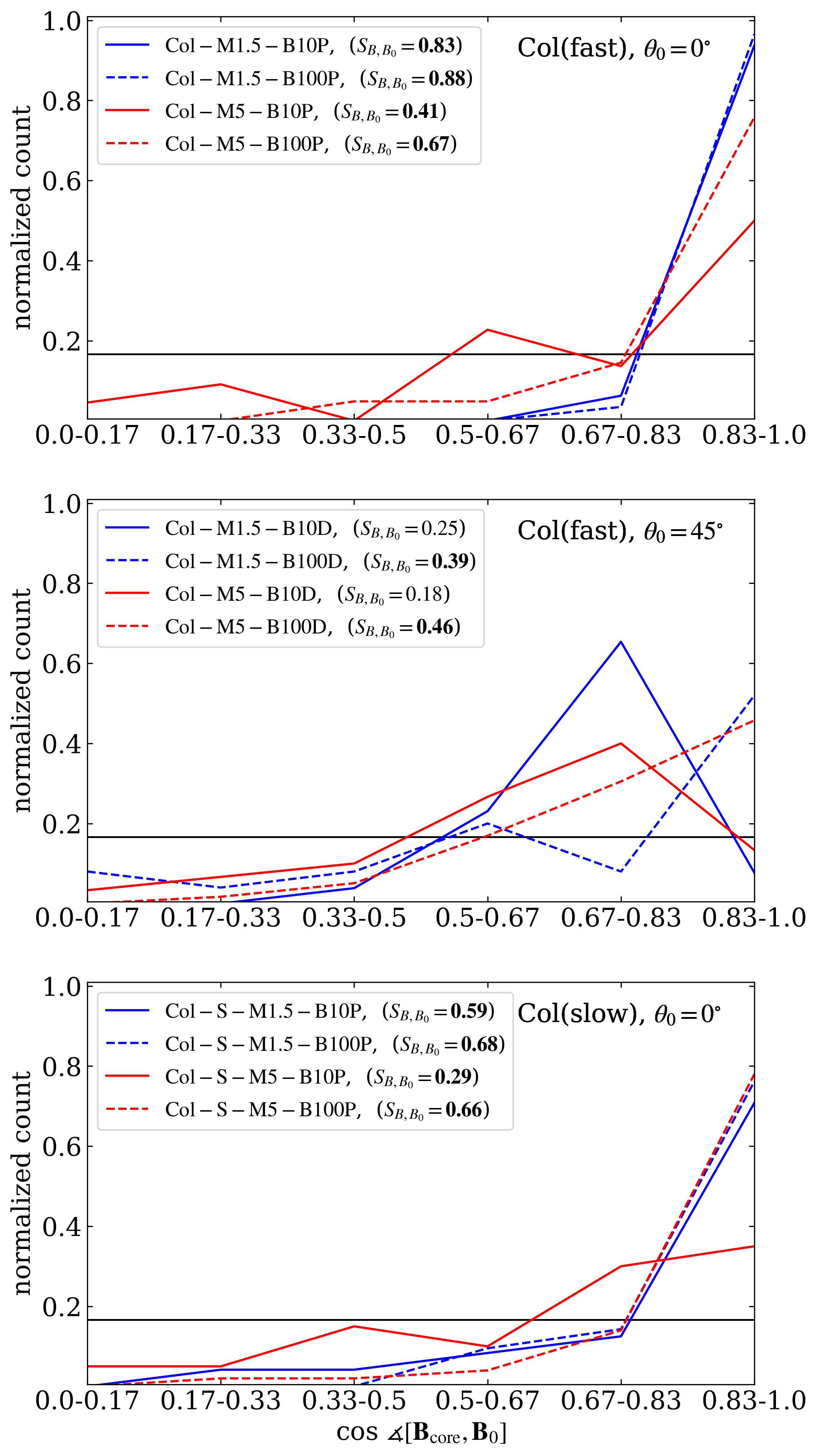}
\end{center}
\caption{Same as Figure \ref{fig:rot_B_B} except for Collision Setup models. Fast collision velocity cases with $\theta_{0}=0^{\circ}$ are shown in the top panel and those of $\theta_{0}=45^{\circ}$ are shown in the middle panel. Slow collision velocity cases are shown in the bottom panel.}
\label{fig:col_B_B}
\end{figure}

\begin{figure}
\begin{center}
  \includegraphics[width=7cm]{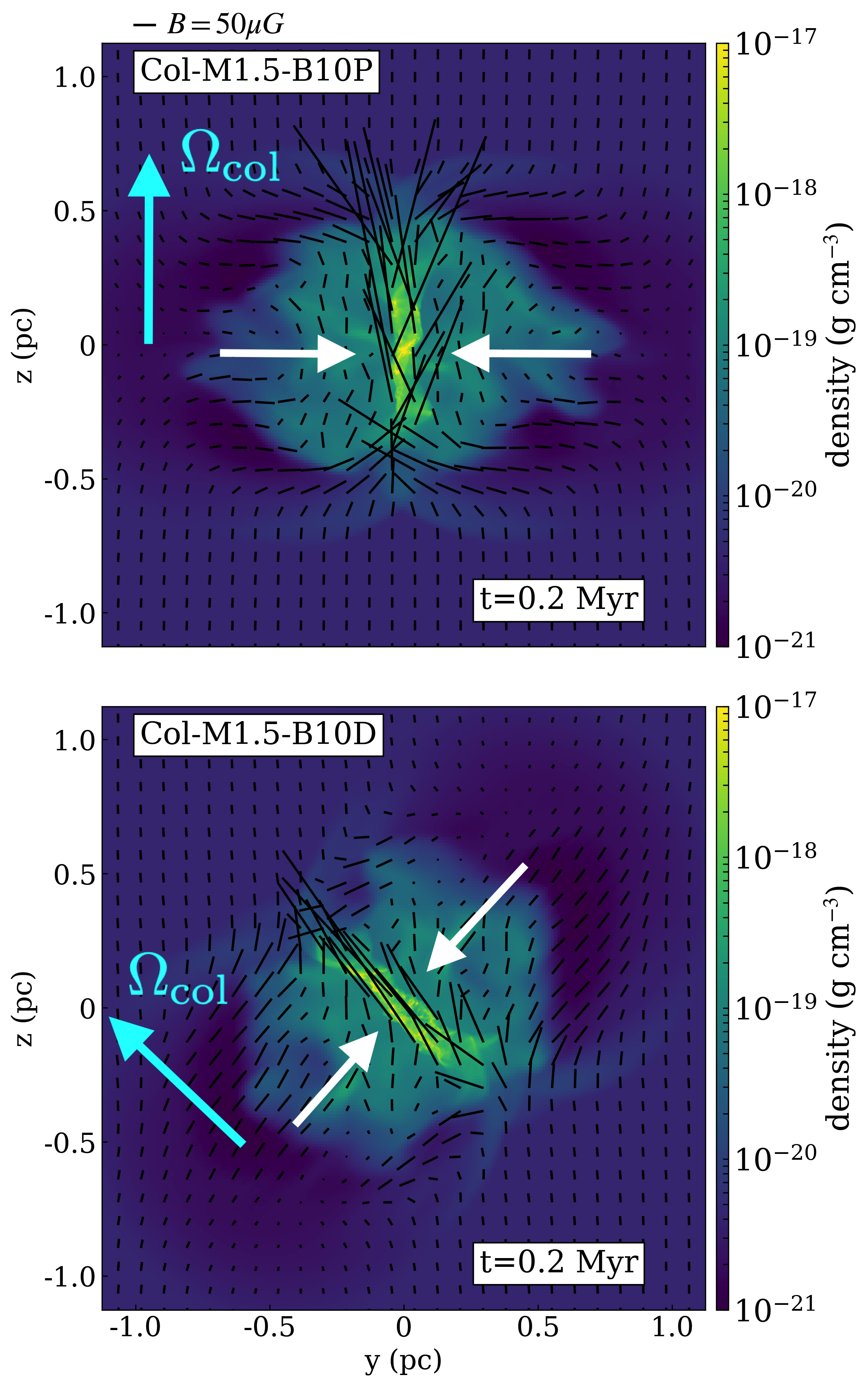}
\end{center}
\caption{Slice plots of the gas density in a plane through the center of the simulation box at $t=0.2~\rm Myr$ in {\tt Col-M1.5-B10P} and {\tt Col-M1.5-B10D} models. Black lines show the magnetic field direction ($\bm{B_{0}}$ is parallel to the $z$ axis). The length of lines corresponds to the magnetic field strength. White arrows indicate the collision axis. The magnetic field is amplified and aligned along the shocked layer. }
\label{fig:col_B_align}
\end{figure}

\begin{figure}
\begin{center}
  \includegraphics[width=8.5cm]{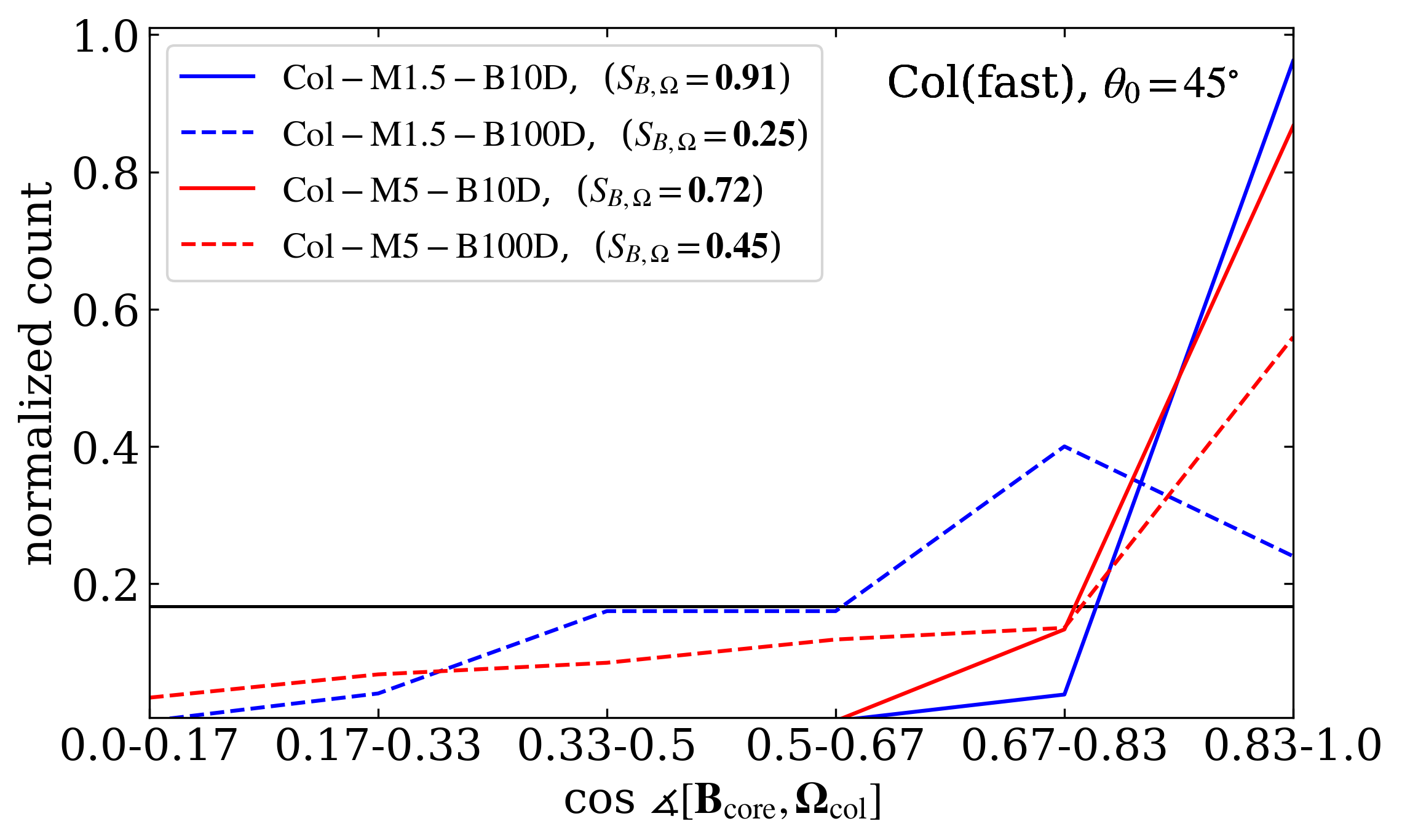}
\end{center}
\caption{Histograms of the cosine of the relative angle between $\bm{B}_{\rm core}$ and $\bm{\Omega}_{\rm col}$ for fast collision models with $\theta_{0}=45^{\circ}$. The black line shows the expected distribution for an isotropic orientation of $\bm{B}_{\rm core}$. In the legends, the orientation parameter $S_{B,\Omega}=(3\langle \mathrm{cos}^{2} \measuredangle[\bm{B}_{\rm core},\bm{\Omega}_{\rm col}]\rangle-1)/2$ for each model are shown, and values with $S_{L,B_{0}}>0.25$ are indicated in bold. 
In models with weak magnetic fields (indicated by solid lines), the strong alignment of $\bm{B}_{\rm core}$ with $\bm{\Omega}_{\rm col}$ is observed. Furthermore, in the weak magnetic field model, $S_{B, \Omega}$ is higher than $S_{B,B_{0}}$, indicating that the direction of $\bm{B}_{\rm core}$ is determined not by the direction of the initial magnetic field, but rather by the direction of collision axis.}
\label{fig:col_B_shock}
\end{figure}

Figure \ref{fig:col_B_B} shows the histograms of cosine of the angle between $\bm{B}_{\rm core}$ and $\bm{B}_{0}$ as Figure \ref{fig:rot_B_B}. In fast collision models with $\theta_{0}=0^{\circ}$ (shown in the top panel), the alignment tendency between $\bm{B}_{\rm core}$ and $\bm{B}_{0}$ is strong, as indicated by a significant deviation from the uniform distribution of $\mathrm{cos} \measuredangle[\bm{B}_{\rm core},\bm{B}_{0}]$, and a large $S_{B,B_{0}}$, which holds true for both strong and weak ${B}_{0}$ models. In Section \ref{sec:rot_Magnetic field}, we showed that in Rotation Setup models with the weak ${B}_{0}$, the direction of $\bm{B}_{\rm core}$ tends to be relatively random, and the degree of alignment between $\bm{B}_{\rm core}$ and $\bm{B}_{0}$ is lower. However, in the Collision Setup, a strong alignment tendency is present even in weak ${B}_{0}$ models. This strong alignment is caused by the large-scale alignment of magnetic fields due to collisions. Figure \ref{fig:col_B_align} is a density and magnetic field (black lines) in slices cut through the center of the simulation box in Collision Setup models with weak $B_{0}$ models, {\tt Col-M1.5-B10P} and {\tt Col-M1.5-B10D}. The shocked layer is formed at the interface of the two clumps. The magnetic field is amplified and aligned more efficiently along the shocked layer as a result of compression and bending by collisions. Most dense cores are formed within the shocked layer, inheriting this aligned magnetic field. Therefore, the degree of alignment of $\bm{B}_{\rm core}$ is high, and this trend is also evident in weak $B_{0}$ models where the magnetic field is easily bent. That is to say, in cases where clumps collide, the direction of $\bm{B}_{\rm core}$ is determined by the direction of the collision axis.

The middle panel of Figure \ref{fig:col_B_B} shows results of fast collision models with $\theta_{0}=45^{\circ}$. In these models with weak ${B}_{0}$ (indicated by solid lines), the peak of the distribution of $\mathrm{cos}\,\measuredangle[\bm{B}_{\rm core},\bm{B}_{0}]$ is not within the range of 0.83-1.0, but rather in the range of 0.67-0.83. Compared to the weak ${B}_{0}$ model with $\theta_{0}=0^{\circ}$ in the top panel, the difference is evident, with smaller values of $S_{B,B_{0}}$. This tendency for misalignment in the $\theta_{0}=45^{\circ}$ model can be explained by the distortion of the magnetic field due to collisions. The bottom panel of Figure \ref{fig:col_B_align} indicates the 
density and magnetic field in slices for the weak ${B}_{0}$ model with $\theta_{0}=45^{\circ}$. Since the collision axis is inclined at $45^{\circ}$ to $\bm{B}_{0}$, the shocked layer is oblique to $\bm{B}_{0}$ and parallel to $\bm{\Omega}_{\rm col}$. The magnetic field inside the clump is twisted and amplified along the direction of the shock, causing the field to be tilted by $45^{\circ}$ with respect to $\bm{B}_{0}$ and become parallel to $\bm{\Omega}_{\rm col}$. $\bm{B}_{\rm core}$ aligns with $\bm{\Omega}_{\rm col}$ rather than $\bm{B}_{0}$, as it inherits the aligned magnetic field within the shocked layer. Figure \ref{fig:col_B_shock} shows histograms of the cosine of the relative angle between $\bm{B}_{\rm core}$ and $\bm{\Omega}_{\rm col}$ for fast collision models with $\theta_{0}=45^{\circ}$. For weak ${B}_{0}$ models, the peak of the distribution of $\mathrm{cos} \measuredangle[\bm{B}_{\rm core},\bm{\Omega}_{\rm col}]$ is sharp within the range of 0.83-1.0 indicating the strong alignment of $\bm{B}_{\rm core}$ with $\bm{\Omega}_{\rm col}$. As $\theta_{0}=0^{\circ}$
models, the collision-axis determines the direction of $\bm{B}_{\rm core}$. However, it should be noted that in cases where ${B}_{0}$ is strong, this may not necessarily hold true. In models with strong ${B}_{0}$ (dashed line in Figure \ref{fig:col_B_shock}), the peak in the distribution is not as sharp compared to the model with weak ${B}_{0}$, and the orientation parameter $S_{B,\Omega}=(3\langle \mathrm{cos}^{2} \measuredangle[\bm{B}_{\rm core},\bm{\Omega}_{\rm col}]\rangle-1)/2$, is smaller. When ${B}_{0}$ is strong, the magnetic field inside the clump is more likely to be aligned with the direction of $\bm{B}_{0}$. Therefore, when $\theta_{0} = 45^{\circ}$, the magnetic field may not align perfectly with the shocked layer. $\bm{B}_{\rm core}$, which inherits the magnetic field inside the clump, will tilt relative to $\bm{\Omega}_{\rm col}$.

\begin{figure}
\begin{center}
  \includegraphics[width=7.0cm]{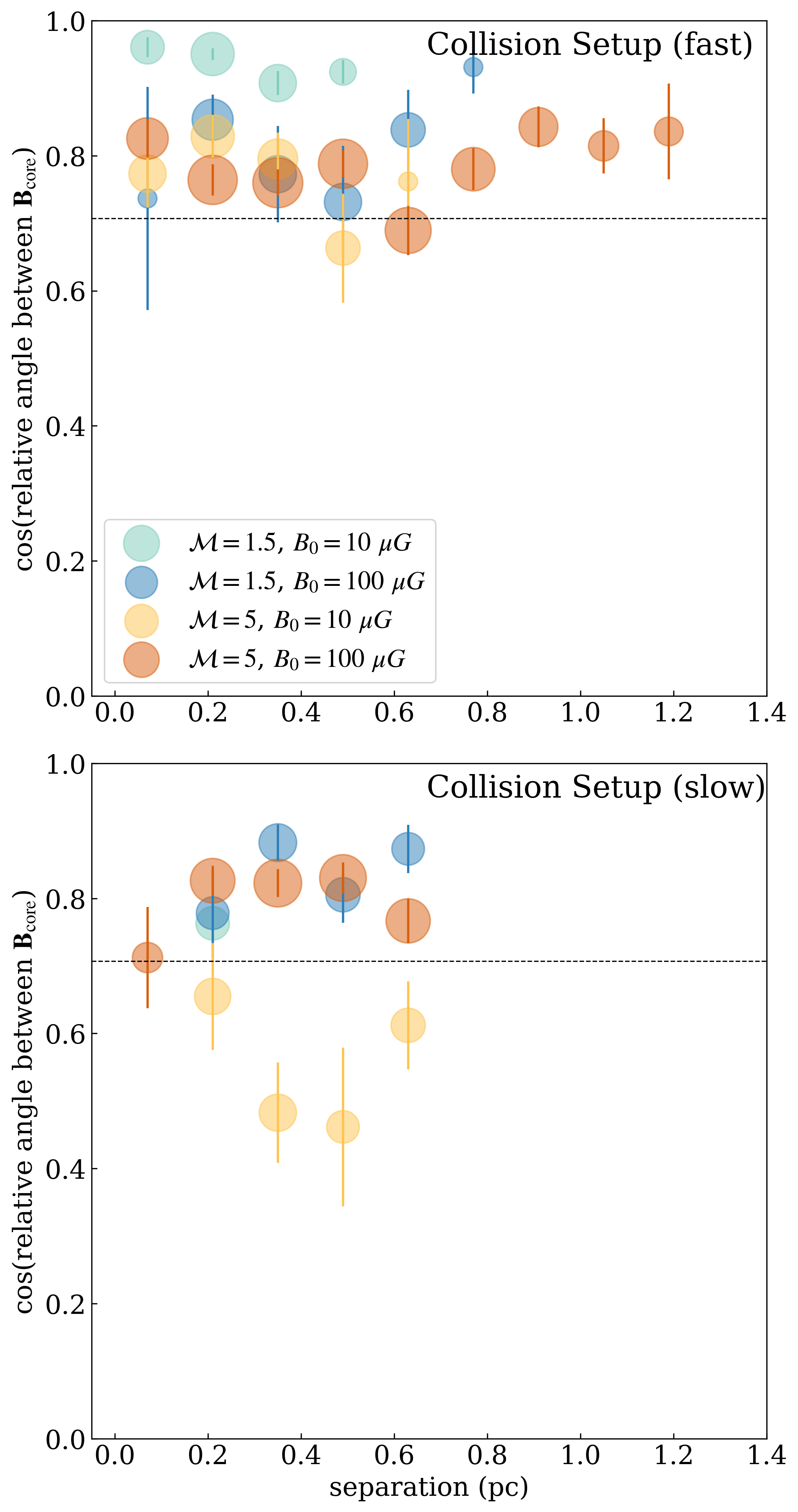}
\end{center}
\caption{Same as Figure \ref{fig:rot_B_pair} except for Collision Setup models. Fast collision velocity cases are shown in the top panel, and slow collision velocity cases are shown in the bottom panel. The Collision Setup model has a higher degree of alignment for $\bm{B}_{\rm core}$ compared to Rotation Setup and w/o Setup models. Especially when $B_{0}$ is weak, the field aligns along the shocked layer, resulting in a significantly higher degree of alignment for $\bm{B}_{\rm core}$.}
\label{fig:col_B_pair}
\end{figure}

Figure \ref{fig:col_B_pair} shows the cosine of the relative orientation angle, noted
as $\mathrm{cos}\,\theta_{B,B}$, between all $\bm{B}_{\rm core}$ pairs as a function of their separation distances in the same manner as Figure \ref{fig:rot_B_pair}. 
All models except for the slow collision model with $\mathcal{M}=5$ and $B_{0}=10\mu G$ exhibit a clear tendency for alignment between $\bm{B}_{\rm core}$ pairs over a wide range of separations. A crucial characteristic is that even in weak $B_{0}$ models, $\mathrm{cos}\,\theta_{B,B}$ is large. As shown in Section \ref{sec:rot_Magnetic field}, in the Rotation setup, $\mathrm{cos}\,\theta_{B,B}$ for weak $B_{0}$ models are significantly smaller than those of strong $B_{0}$ models. However, in the Collision setup, the global alignment of the fields inside the clump due to collisions causes $\mathrm{cos}\,\theta_{B,B}$ of the weak $B_{0}$ models to be even greater.
In the slow collision model with $\mathcal{M}=5$ and $B_{0}=10\mu G$, due to the strong turbulence and slow collision velocities, the magnetic field is not well aligned at the shocked layer, resulting in a random distribution of the direction of $\bm{B}_{\rm core}$. However, in other models, if collisions compress the gas sufficiently, the $\bm{B}_{\rm core}$ pairs will align. Generally, the orientation of the collision axis (or $\bm{\Omega}_{\rm col}$) is a crucial factor in determining the direction of $\bm{B}_{\rm core}$.

\subsubsection{Rotation-Magnetic field relation of the Collision Setup}
\label{sec:col_Rotation-Magnetic field relation}

\begin{figure}
\begin{center}
  \includegraphics[width=5.85cm]{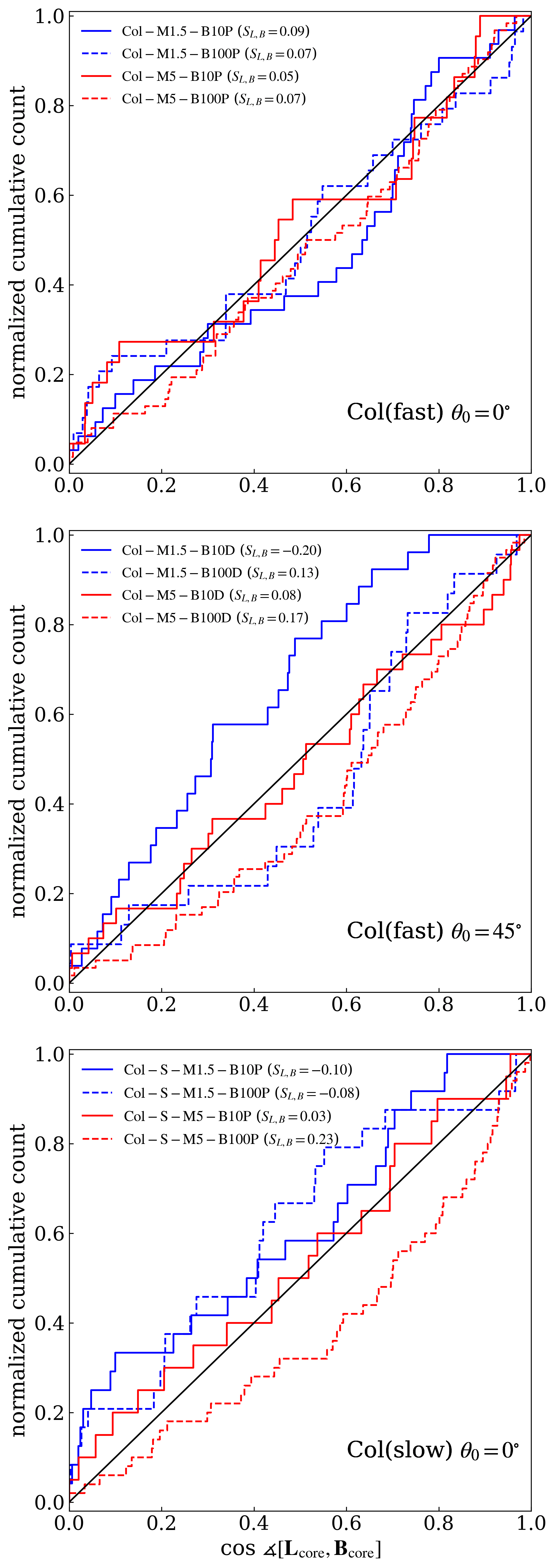}
\end{center}
\caption{Same as Figure \ref{fig:rot_LB} except for Collision Setup models. Fast collision velocity cases with $\theta_{0}=0^{\circ}$ are shown in the top panel and those of $\theta_{0}=45^{\circ}$ are shown in the middle panel. Slow collision velocity cases are shown in the bottom panel. In most models, they suggest random distributions of $\measuredangle[\bm{L}_{\rm core},\bm{B}_{\rm core}]$.}
\label{fig:col_LB}
\end{figure}

Figure \ref{fig:col_LB} illustrates the cumulative distribution function (CDF) of the cosine of the relative angle between $\bm{L}_{\rm core}$ and $\bm{B}_{\rm core}$ compared to a uniform distribution as Figure \ref{fig:rot_LB}. For most models, CDF is similar to a uniform distribution, and the null hypothesis that "the distribution is uniform" cannot be rejected at a significance level of 5\% using the K-S test, except for models {\tt Col-M5-B100D} and {\tt Col-S-M5-B100P}.
The models {\tt Col-M5-B100D} and {\tt Col-S-M5-B100P} with strong $B_{0}$ show a slight tendency towards alignment, which may be attributed to the magnetic braking by fields that are intensified by the collision. However, $S_{L,B}$ is not significantly large (weak alignment), and other models of Collision Setup show almost uniform distributions.

\section{Discussion}

% energy ration (table)
% virial parameter 

\label{sec:Discussion}
\subsection{Alignment of Core Angular Momentum}
\label{dis:Alignment of Core Angular Momentum}
The inheritance of global motion by dense cores from their parental clump is a significant topic of study. Previous research has yielded mixed results, with some studies indicating mostly random distributions of spin axes in young open clusters \citep{10.1111/j.1365-2966.2009.15983.x}, while others have found strong spin alignment of stars within specific open clusters \citep{Corsaro_2017}. It is likely that this variability depends on the specific environment of the parental star-forming regions. In this subsection, we discuss the trend of the relative angle between the parental clump rotation axis and $\bm{L}_{\rm core}$ as shown in Section \ref{sec:rot_Angular momentum} and \ref{sec:col_Angular momentum}.

\begin{figure}
\begin{center}
  \includegraphics[width=6.5cm]{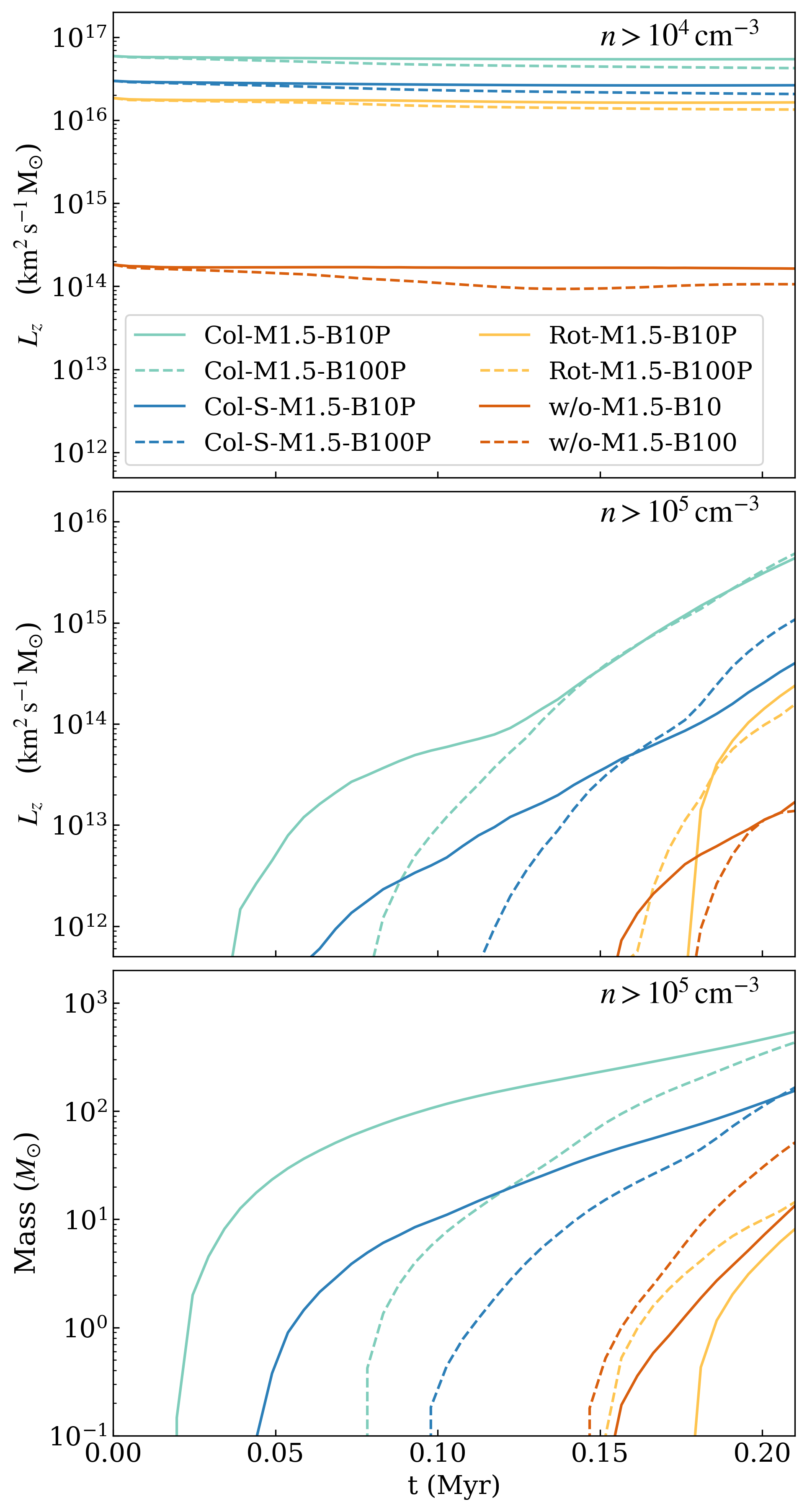}
\end{center}
\caption{Total angular momentum of clump gas in the $z$-direction, $L_{\rm z}$, as a function of time. The top and middle panels display the time evolution of gas above density $10^{4}$ and $10^{5}\,\rm cm^{-3}$, respectively. The bottom panel illustrates the time evolution of the total gas mass above density $10^{5}\,\rm cm^{-3}$ for comparison. We are comparing models for weak turbulence $\mathcal{M}=1.5$ and $\theta_{0}=0$, where $\bm{\Omega}_{0}$ ($\bm{\Omega}_{\rm col}$) is parallel to the $z$-axis. Since the angular momentum of the clump is approximately conserved, $L_{\rm z}$ with $n>10^{4}\,\mathrm{cm}^{-3}$ is nearly constant. On the other hand, $L_{\rm z}$ with $n>10^{5}\,\mathrm{cm}^{-3} (>n_{\rm clump}=1.2\times10^{4}\,\mathrm{cm}^{-3} )$ increases as dense regions formed.}
\label{fig:L_z_time}
\end{figure}

Figure \ref{fig:L_z_time} shows the time evolution of the total angular momentum of clump gas in the $z$-direction, $L_{\rm z}$ for $\mathcal{M}=1.5$ and $\theta_{0}=0^{\circ}$ models. The top panel indicates the $L_{\rm z}$ of gas with number density $n>10^4\,\mathrm{cm}^{-3}$. Since the angular momentum of the clump is approximately conserved, $L_{\rm z}$ remains nearly constant. The middle panel illustrates the $L_{\rm z}$ of gas with $n>10^5\mathrm{cm}^{-3}(>n_{\rm clump}=1.2\times10^{4}\,\mathrm{cm}^{-3} )$. In the models of Rotation Setup and Collision Setup, dense gas inherits the overall rotation of clumps, resulting in higher $L_{\rm z}$ than w/o Setup models.

As shown in Section \ref{sec:rot_Angular momentum} and \ref{sec:col_Angular momentum}, for all strong turbulence models ($\mathcal{M}=5$), we find the random distributions of $\measuredangle[\bm{L}_{\rm core},\bm{\Omega}_{0}(\bm{\Omega}_{\rm col})]$. Strong turbulence can disturb the gas feeding the core, causing a loss of memory of the clump's global rotational motion. In the Rotation Setup with weak turbulence ($\mathcal{M}=1.5$), we observed the alignment between $\bm{\Omega}_{0}$ and $\bm{L}_{\rm core}$. Cores inherit the angular momentum of dense regions indicated in Figure \ref{fig:L_z_time}.
However, in the Collision Setup, even with weak turbulence, the distributions of $\mathrm{cos} \measuredangle[\bm{L}_{\rm core},\bm{\Omega}_{\rm col}]$ are uniform, and the $\bm{L}_{\rm core}$ pairs are not aligned with each other. The direction of $\bm{L}_{\rm core}$ does not reflect the larger angular momentum of clumps illustrated in Figure \ref{fig:L_z_time}.

\begin{figure}
\begin{center}
  \includegraphics[width=6.5cm]{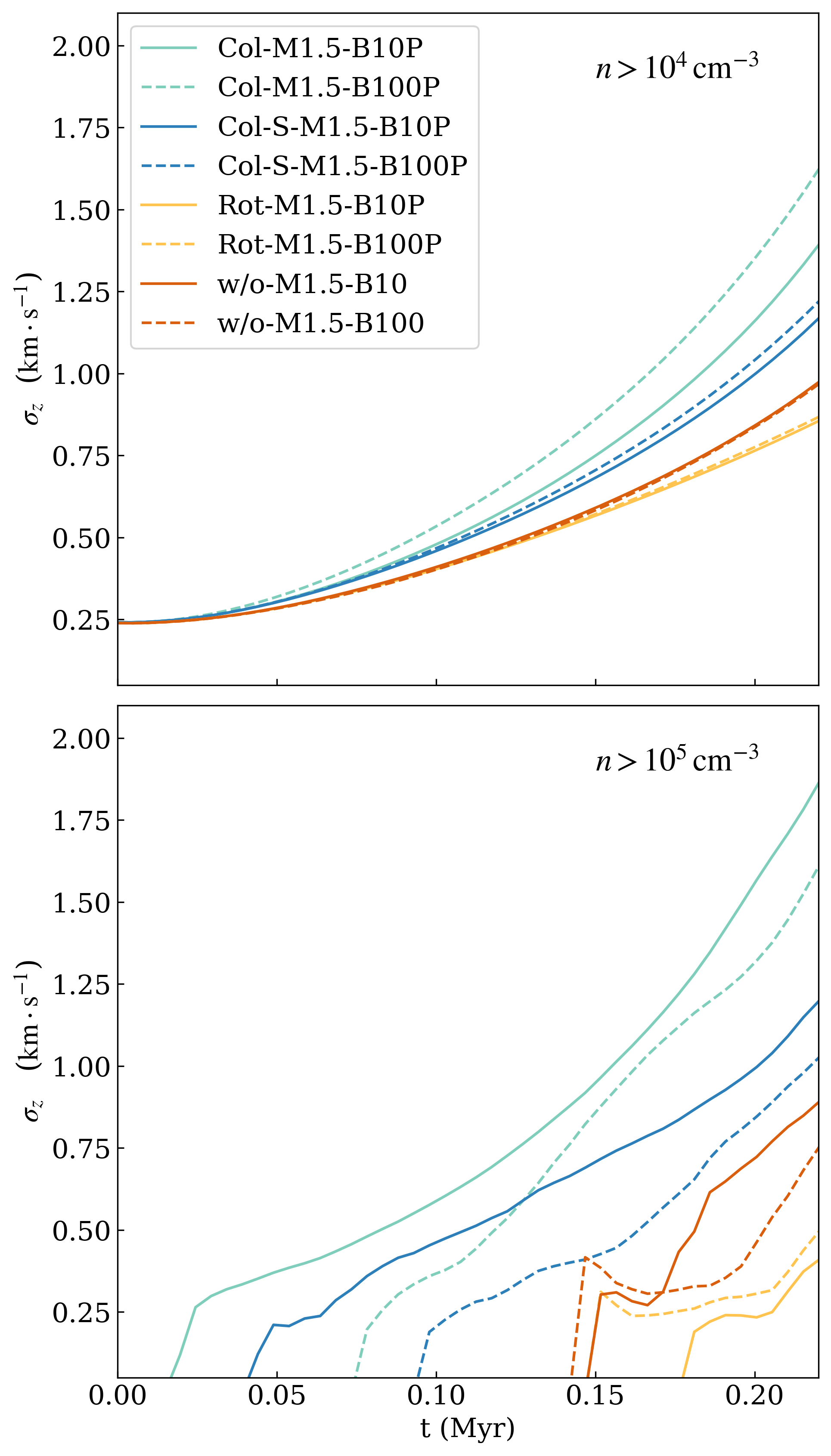}
\end{center}
\caption{Velocity dispersion in $z$-direction, $\sigma_{\rm z}$, as a function of time. The top and bottom panels display the time evolution of gas above density $10^{4}$ and $10^{5}\,\rm cm^{-3}$, respectively. Increasing the initial collision velocity leads to a higher increase in $\sigma_{\rm z}$.}
\label{fig:sigma_v}
\end{figure}

\begin{figure*}
\begin{center}
  \includegraphics[width=12cm]{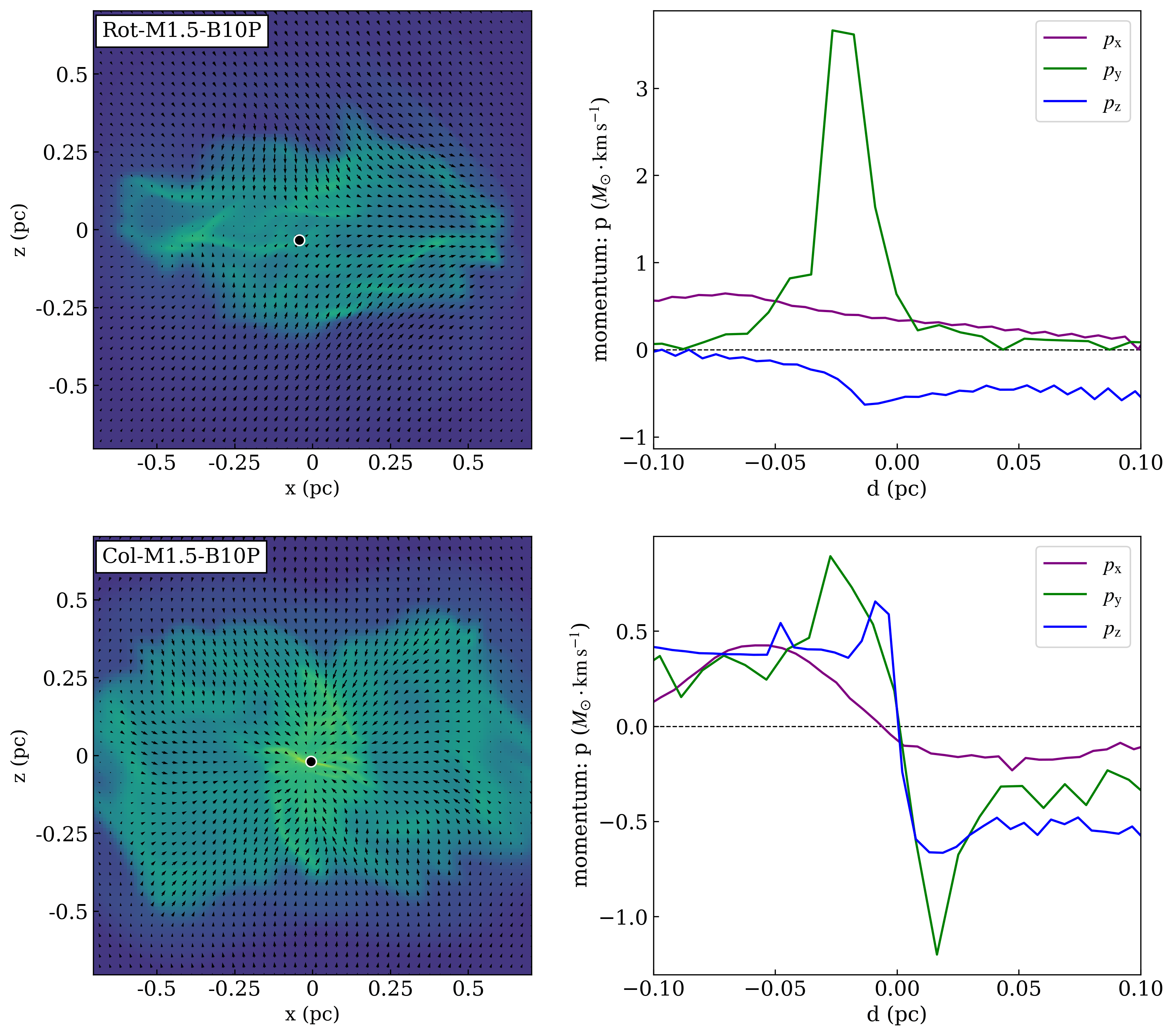}
\end{center}
\caption{Left: Column densities summed over 0.6 pc for the {\tt Rot-M1.5-B10P} model at 0.3 Myr (top) and {\tt Col-M1.5-B10P} model at 0.2 Myr (bottom). The black lines show the average velocity field projected onto the plane. The position of the most massive dense core is indicated with a black circle. Right: Momentum structure around the most massive dense core corresponding to the left panel. Purple lines show the average momentum around the $x$-axis: $\int_S \rho (v_x-v_{x \rm ,core}) d y d z / \int_S d y d z$, where $v_{x \rm ,core}$ is the $x$-component of the center of mass velocity of the core, and $S=\{(y, z) \mid\left[\left(y-y_{\text {core }}\right)^2+\left(z-z_{\text {core }}\right)^2\right]^{1 / 2} \leq 0.05 \mathrm{pc}\}$. The horizontal axis corresponds to $x-x_{\text {core }}$. Similarly, the green and blue lines represent the average momentum around the $y$-axis and $z$-axis, respectively.
Momentum around the core is isotropic in the {\tt Col-M1.5-B10P} model, while the $y$-component is dominant in the {\tt Rot-M1.5-B10P} model. }
\label{fig:core_rotation}
\end{figure*}

One of the factors contributing to this misalignment is the turbulence induced by the collision. Figure \ref{fig:sigma_v} follows the time evolution of velocity dispersion in $z$-direction, $\sigma_{\rm z}$ for $\mathcal{M}=1.5$ and $\theta_{0}=0^{\circ}$ models. In Collision Setup models with $\theta_{0}=0^{\circ}$, $\sigma_{\rm z}$ can be a measure of the driven turbulence by collision because it is a component perpendicular to the direction of the collision axis and following rotation velocity. For two density thresholds, the Collision Setup models produce higher $\sigma_{\rm z}$ than other setup models
roughly by a factor of two. Also, models with higher collision velocities (fast) drive more turbulence. The increase in turbulence intensity due to collisions has also been demonstrated in \citep{10.1093/pasj/psx140}, which is consistent with our results. Dense shocked regions formed by collision have greater velocity dispersion perpendicular to the collision axis (parallel to $\bm{\Omega}_{\rm col}$). Therefore, the motion of gas around the dense core is more turbulent, and the orientation of the core rotation axis is considered to be random without any specific direction.

In Figure \ref{fig:core_rotation}, to show the direction dependency of the accretion flow toward the dense core, we illustrate the momentum structure around the most massive dense core position for two models, {\tt Rot-M1.5-B10P} of Rotation Setup and {\tt Col-M1.5-B10P} of Collision Setup. In the left panels, we display Column densities summed over 0.6 pc around the most massive core. In the right panels, the purple lines show the average momentum around the $x$-axis: $\int_S \rho (v_x-v_{x \rm, core}) d y d z / \int_S d y d z$, where $v_{x \rm, core}$ is the $x$-component of the center of mass velocity of the core, and $S=\{(y, z) \mid\left[\left(y-y_{\text {core }}\right)^2+\left(z-z_{\text {core }}\right)^2\right]^{1 / 2} \leq 0.05 \mathrm{pc}\}$. The horizontal axis corresponds to $x-x_{\text {core }}$. Similarly, the green and blue lines represent the average momentum around the $y$-axis and $z$-axis, respectively. In the Rotation Setup model {\tt Rot-M1.5-B10P}, gas with predominant momentum in the $y$ direction can be observed around the core. We have confirmed that, similarly, gas with dominant momentum in the $x$ or $y$ direction can be found around many other cores. In general, in Rotation Setup models, the velocity field of clump rotation is parallel to the $xy$ plane. Therefore, dense gas regions with dominant momentum in the $xy$ direction are more likely to form, especially when turbulence is weak. As momentum in such a region is injected into the core, the dense core acquires a rotational velocity on the $xy$ plane, with the rotation axis parallel to the $z$-axis. The same effect occurs even when $\theta_{0}=45^{\circ}$. Thus, as shown in Section \ref{sec:rot_Angular momentum}, in weak turbulence Rotation Setup, $\bm{L}_{\rm core}$ aligns with $\bm{\Omega}_{0}$.

On the other hand, the bottom-right panel of Figure \ref{fig:core_rotation} shows that the momentum of the gas around the core is isotropic in the Collision Setup model {\tt Col-M1.5-B10P}. Momentum is not biased in the direction of clump motion ($xy$ direction), and gas flow from the direction perpendicular to the collision axis ($z$) is also injected into the core to a significant degree. A high-density shocked layer is formed perpendicular to the collision axis, so the flow along the shocked layer (parallel to the $z$-axis) significantly contributes to the momentum supply to the core. We observed that many other cores in other Collision Setup models also exhibit an isotropic tendency toward accretion. Therefore, in Collision Setup, the rotation direction of the dense core is not biased to one side, and the distributions of $\mathrm{cos} \measuredangle[\bm{L}_{\rm core},\bm{\Omega}_{\rm col}]$ are uniform, as shown in Section \ref{sec:col_Angular momentum}. 

\subsection{Misalignment between the angular momentum and the magnetic field}

\citet{10.1111/j.1745-3933.2010.00942.x} showed that the efficiency of mass ejection in the outflow is dependent on the angle between the rotation axis and the magnetic field. Therefore, considering the launching of outflows, the relationship between rotation and magnetic fields is crucial. The classical theory suggests that the magnetic axis of cores should be parallel to their rotational axis, as perpendicular configuration allows for faster magnetic braking compared to parallel configurations \citep{1979ApJ...228..159M, Mouschovias_1979}. However, \citet{Hull_2013} and \citet[and references therein]{2019FrASS...6....3H} showed the random orientations of the core rotation and magnetic fields within protostellar cores in various regions in the whole sky. \citet{Doi_2020} also find no correlation between the magnetic field angles and the rotation axes in NGC 1333. On the other hand,
there are observations suggesting the preference for weak alignment of them in some regions \citep{2021ApJ...907...33Y,Xu_2022}. \citet{Kong_2019} indicated outflow axes being mostly orthogonal to their parent filament in G28.37+0.07, which may suggest a preferred alignment between the core rotation and bulk fields.

\begin{figure*}
\begin{center}
  \includegraphics[width=17cm]{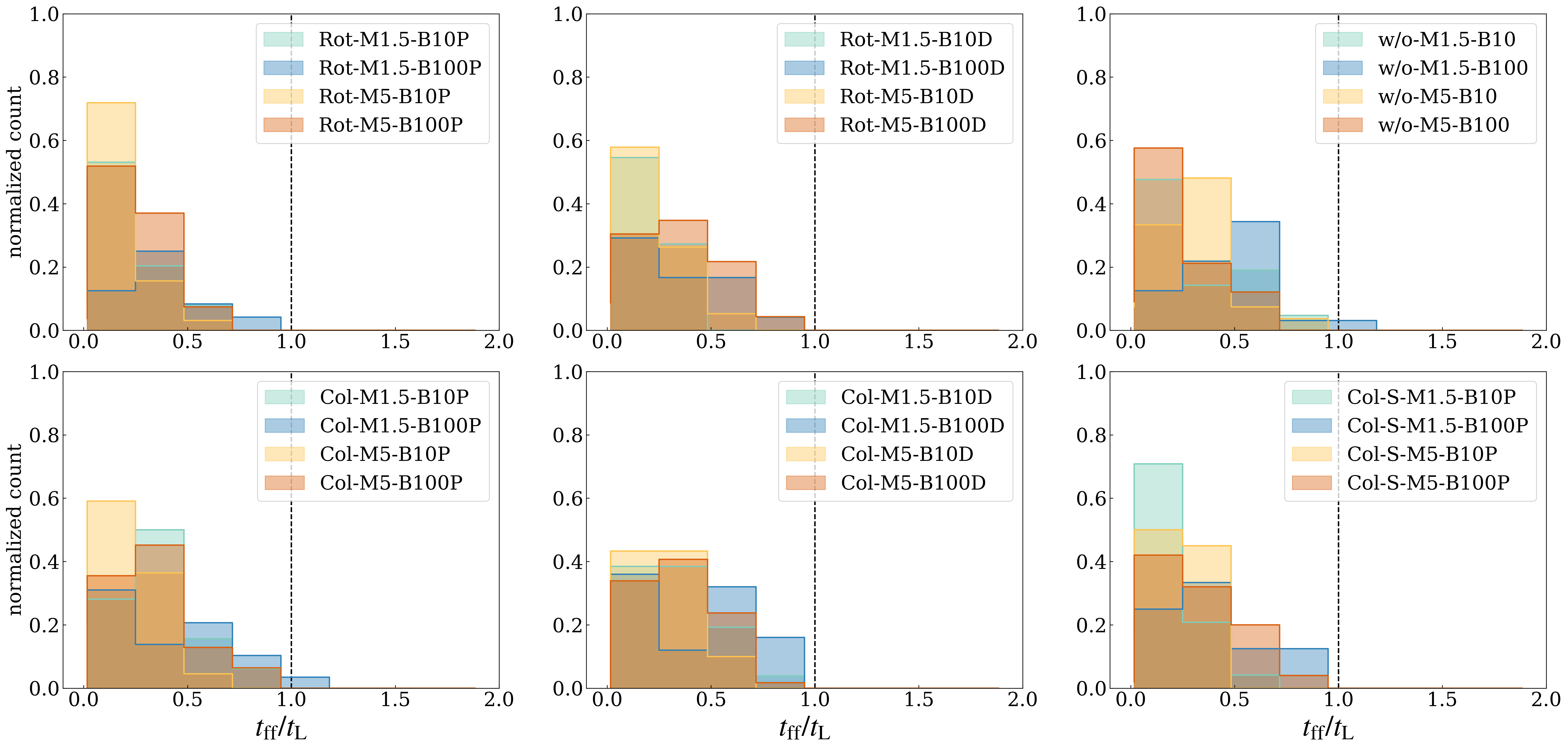}
\end{center}
\caption{Histograms of the ratio of the free fall time $t_{\rm ff}$ to the magnetic braking timescale $t_{\rm L}\equiv \sigma/2\rho_{\rm ext}v_{\rm A}$ for all bound cores in different models. Vertical dashed lines indicate the $t_{\rm ff}/t_{\rm L}=1.0$. For most cores, $t_{\rm ff}/t_{\rm L}$ is lower than 1.0, suggesting that the effect of magnetic braking is not large.}
\label{fig:t_L}
\end{figure*}

As shown in Section \ref{sec:rot_Rotation-Magnetic field relation} and \ref{sec:col_Rotation-Magnetic field relation}, we find that the relative angle between $\bm{L}_{\rm core}$ and $\bm{B}_{\rm core}$ is random in most models. The alignment is strong only for the model {\tt Rot-M1.5-B100P} due to the limited initial conditions. In some models of the Collision Setup, random distributions are rejected, but the alignment is weak.
These results of general misalignment are consistent with previous numerical simulations \citep{2016ApJ...827L..11O, Chen_2018,Kuznetsova_2020}. Our findings suggest that, except for special cases, there is no tendency for strong alignment between $\bm{L}_{\rm core}$ and $\bm{B}_{\rm core}$ at core-scale ($\sim 0.01-0.1\mathrm{pc})$ regardless of the clump properties. This suggests that the effect of magnetic braking to align $\bm{L}_{\rm core}$ and $\bm{B}_{\rm core}$ is weak.

We use the characteristic time $t_{\rm L}$ for magnetic braking to study the effect of magnetic braking quantitatively. When the mean direction of the field lines is parallel to the rotation axis, $t_{\rm L}$ is given by

\begin{equation}
\label{eq:t_L}
t_{\rm L}\equiv\frac{1}{2}\frac{\sigma}{\rho_{\rm ext}v_{\rm A}},
\end{equation}
where $\sigma$ is the column density of the core, $\rho_{\rm ext}$ and $v_{\rm A}$ are the density and the Alfvén velocity in the ambient medium \citep{Mouschovias_1980}. $t_{\rm L}$ is an approximate timescale for magnetic braking to constrain the angular momentum. 
We calculate $\sigma$ as $\sigma=M_{\rm core}/\pi R_{\rm core}^{2}$, $\rho_{\rm ext}$ as the density at the core boundary ($=\mu m_{\rm H}n_{\rm th}$), and $v_{\rm A}$  as $B_{\rm core}/(4\pi\rho_{\rm ext})^{1/2}$. Figure \ref{fig:t_L} shows the histogram of the ratio between the free-fall time $t_{\rm ff}$ and $t_{\rm L}$ for all bound cores in different models. For most cores, $t_{\rm ff}/t_{\rm L}<1$, indicating that the effect of magnetic braking is not significant. This is consistent with the random distribution of $\measuredangle[\bm{L}_{\rm core},\bm{B}_{\rm core}]$ shown in 
Section \ref{sec:rot_Rotation-Magnetic field relation} and \ref{sec:col_Rotation-Magnetic field relation}. Our simulations are ideal MHD, but the magnetic braking can be even weaker if the non-ideal MHD effects (including ambipolar diffusion, Hall effect, and Ohmic dissipation) are considered \citep[e.g.,][]{2009ApJ...698..922M,2010MNRAS.408..322K,2015ApJ...801..117T,2016A&A...587A..32M,2018A&A...619A..37M}. 

We note that $t_{L}$ defined in Equation \ref{eq:t_L} applies to the case where the magnetic field and the angular momentum are parallel. Our identified cores include those in which $\bm{L}_{\rm core}$ is not parallel to $\bm{B}_{\rm core}$, therefore Equation \ref{eq:t_L} is only approximate. The cores we studied are gravitationally bound, and magnetic energy is not dominant. Hence, we cannot rule out the possibility that $\bm{L}_{\rm core}$ aligns with $\bm{B}_{\rm core}$  due to the magnetic braking in cores with lower masses and higher magnetic field effects.

In Appendix \ref{app:LB-energy}, we also show the correlation between $\measuredangle[\bm{L}_{\rm core},\bm{B}_{\rm core}]$ and energies of cores for all models. We found that $\measuredangle[\bm{L}_{\rm core},\bm{B}_{\rm core}]$ is independent of $E_{\rm mag}/|E_{\rm grav}|$ or $E_{\rm kin}/|E_{\rm grav}|$, and their distribution is random for our samples.

\subsection{Dynamics}
\label{sec:Dynamics}

\begin{figure*}
\begin{center}
  \includegraphics[width=16cm]{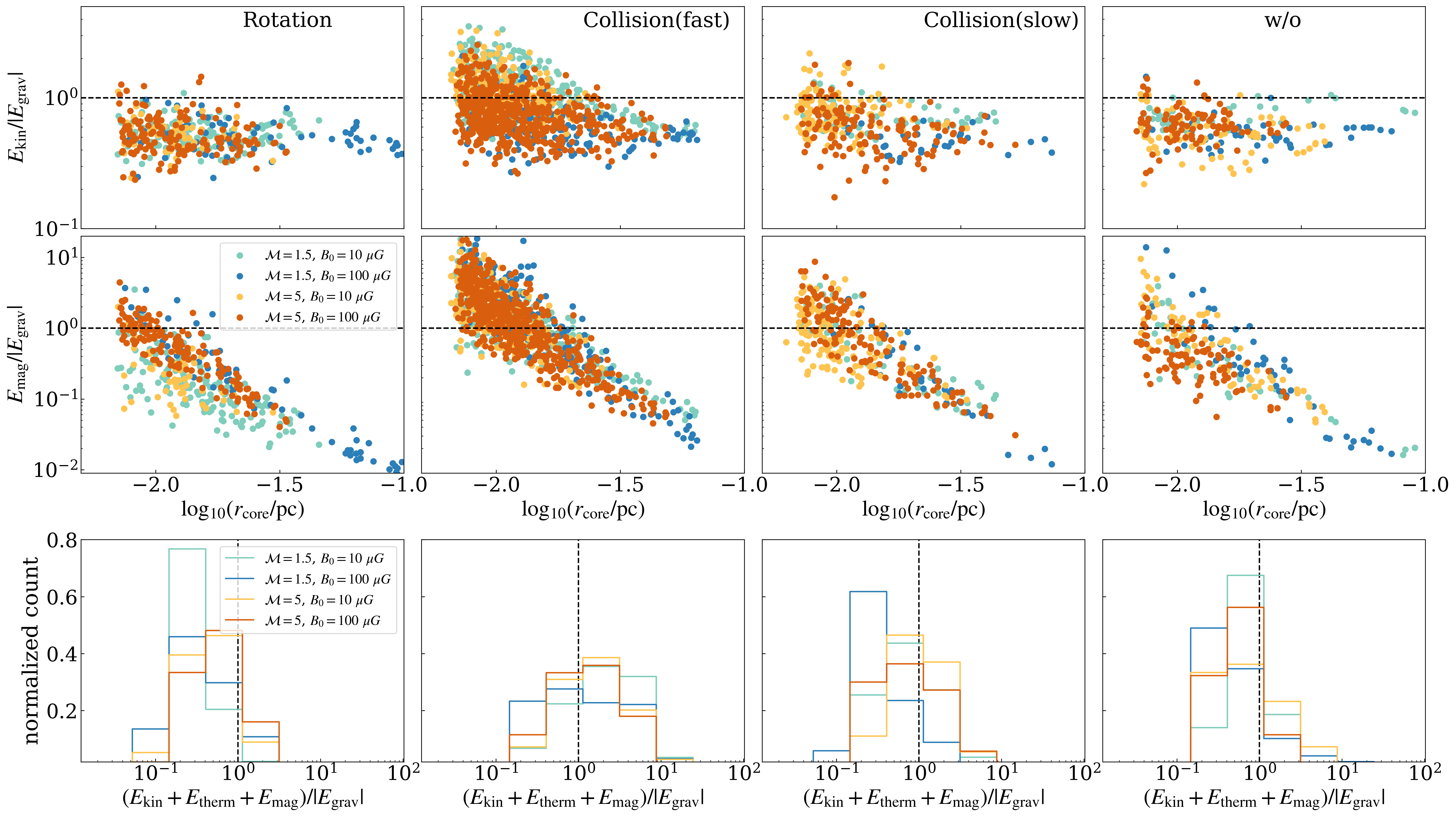}
\end{center}
\caption{Top row: Ratio between kinetic and gravitational energies, $E_{\rm kin}/|E_{\rm grav}|$, as a function of core radius. The models with $\theta_{0}=0^{\circ}$ and $45^{\circ}$ are presented together. Note that, unlike the other figures, this plot includes the unbound ($E_{\rm thermal}+E_{\rm kin}+E_{\rm mag}+E_{\rm grav}>0 $) core. Results of Rotation Setup, Collision Setup(fast), Collision Setup(slow), and w/o Setup are shown from left to right. Middle row: Ratio between magnetic and gravitational energies, $E_{\rm mag}/|E_{\rm grav}|$, plotted against the core radius. Bottom row: Histograms of the energy ratio of the sum of turbulent, thermal field, and magnetic field energies to the absolute value of self-gravitational energy. In the Collision Setup, the magnetic fields and kinetic energy have a relatively larger impact on smaller cores, resulting in a higher ratio of unbound cores.}
\label{fig:energy_unbound}
\end{figure*}

Energy analysis can reveal the dynamical properties of the cores and provide an important indicator in exploring the impact of the clump environment on the core. In this subsection, we mainly discuss energies of dense cores identified in our simulation. 

The first row of Figure \ref{fig:energy_unbound} present the ratio between kinetic and gravitational energies, $E_{\rm kin}/|E_{\rm grav}|$, as a function of core radius. Unlike the other figures, this plot includes unbound cores ($E_{\rm thermal}+E_{\rm kin}+E_{\rm mag}+E_{\rm grav}>0 $). In the Rotation and w/o Setups, $E_{\rm kin}/|E_{\rm grav}|$ is relatively independent of the radius. In the Collision Setup, for smaller cores, the contribution of kinetic energy is relatively large compared to other setups. This is due to the turbulence induced by the collision, which was transferred to the cores (see also Section \ref{dis:Alignment of Core Angular Momentum}). 
These results are consistent with the previous work by \citet{2023MNRAS.522..700H} and support their findings.

In the second row, we present the ratio between magnetic and gravitational energies, $E_{\rm mag}/|E_{\rm grav}|$, indicating that the contribution of $E_{\rm mag}$ is significant in the Collision Setup. As shown in Section \ref{sec:rot_Magnetic field}, the gas is compressed perpendicular to the magnetic field in the Collision Setup, and the magnetic field is amplified inside the core while remaining aligned. This suggests that the magnetic field is strengthened more compared to other models where gas contracts isotropically \citep[e.g.,][]{galaxies9020041}. A common trend in all models is that the contribution of $E_{\rm mag}$ decreases with increasing core radius. In other words, the larger the core, the smaller the contribution of $E_{\rm mag}$ to $E_{\rm grav}$ and $E_{\rm kin}$. In particular, the $E_{\rm mag}$ is not dominant in the bound core. This is qualitatively consistent with the result that $\bm{B}_{\rm core}$ does not limit $\bm{L}_{\rm core}$ as we have shown in Section \ref{sec:Results}. 

In the third row, we show histograms of the energy ratio of the sum of turbulent, thermal, and magnetic field energies to the absolute value of self-gravitational energy. As mentioned above, the contribution of turbulent and magnetic energies is significant in the Collision Setup, resulting in a higher proportion of unbound cores (see also Appendix \ref{app:Energetic properties of cores}). A significant number of unbound cores are also reported by previous observations \citep{2010ApJ...714..680M, 2015MNRAS.450.1094P,2017ApJ...846..144K, 2023ApJS..264...35T} and turbulent simulations of clustered star formation \citep{2011ApJ...740...36N, 2022MNRAS.517..885O}.
They are often confined by surface forces by turbulence and magnetic fields.
In summary, the formation of small-sized cores with a significant contribution of $E_{\rm kin}$ and $E_{\rm mag}$ is a characteristic unique to the Collision Setup.

\begin{figure*}
\begin{center}
  \includegraphics[width=18cm]{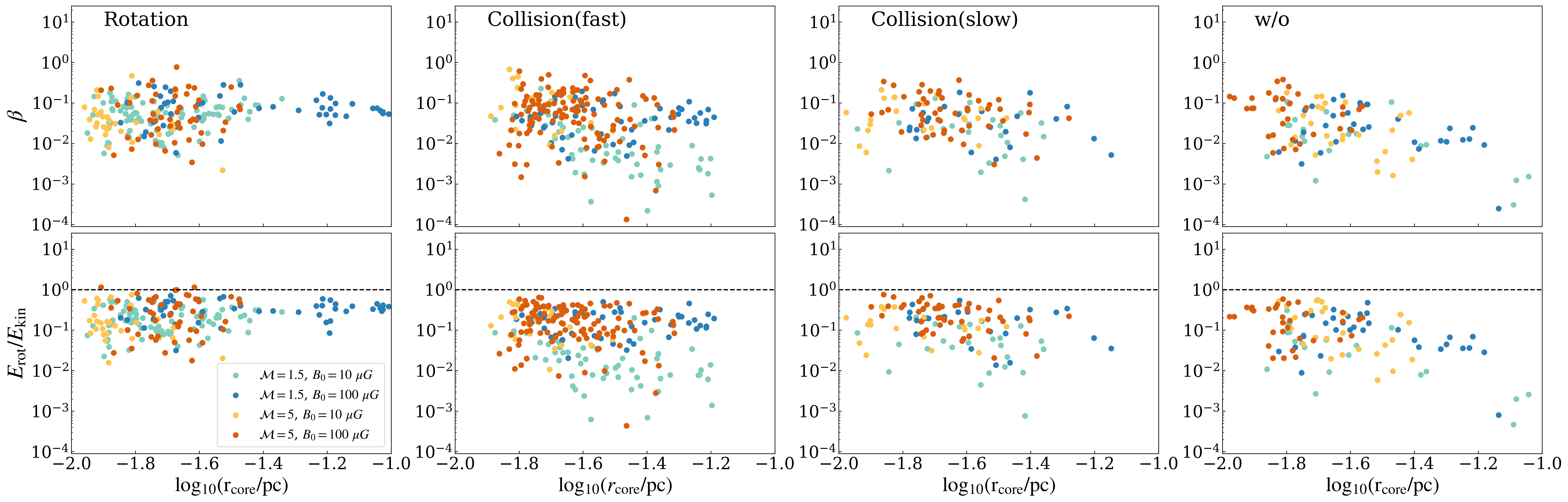}
\end{center}
\caption{Top row: Rotational parameter $\beta$ as a function of bound core radius. The models with $\theta_{0}=0^{\circ}$ and $45^{\circ}$ are presented together. Results of Rotation Setup, Collision Setup(fast), Collision Setup(slow), and w/o Setup are shown from left to right. Bottom row: Ratio between rotational and total kinetic energies, $E_{\rm rot}/E_{\rm kin}$, plotted against the core radius. The distribution of $\beta$ exhibits a large scatter with a typical value of $\beta\sim 0.05$, and there is also a large scatter in the distribution of $E_{\rm rot}/E_{\rm kin}$. The Rotation Setup model tends to have a lower scatter and slightly larger $E_{\rm rot}/E_{\rm kin}$ values than other setups. However, it is concluded that rotation is not the dominant motion in the dense core for any of the models.}
\label{fig:energy}
\end{figure*}

Next, we will consider the contribution of the rotational energy of the bound cores. The ratio between the rotational energy and the gravitational potential energy, $\beta\equiv E_{\rm rot}/|E_{\rm grav}|$, is sometimes referred to as the “rotational parameter”. The first row of Figure \ref{fig:energy} confirms that $\beta$ has a fairly large scatter, and there is no clear dependence on radius. In the Rotation Setup, the scatter of larger cores is relatively small and $\beta$ is high, but as with other models, the typical value is $\beta\sim 0.05$ (see also Appendix \ref{app:property}). This value is generally consistent with observations \citep[e.g.,][]{1993ApJ...406..528G,Caselli_2002,Tobin_2011,Chen_2019}. The second row of Figure \ref{fig:energy} shows the ratio between rotational and kinetic energies, $E_{\rm rot}/E_{\rm kin}$. In Collision and w/o Setups, this ratio exhibits significant variation, with the median ranging from a few percent to several tens of percent. In Rotation Setup, although the variation for larger cores is relatively low, $E_{\rm rot}/E_{\rm kin}$ is at most around several tens of percent, indicating that the contribution of rotation is not significant (see also Table \ref{tab:Properties of identified cores}).
We can conclude that even if the parental clump is rotating or colliding, rotational motion is not dominant within bound dense cores.
However, a characteristic feature of cores in Rotation Setup is that the contribution of $E_{\rm rot}$ is higher than that of other Setups, and the variance of $E_{\rm rot}/E_{\rm kin}$ and $E_{\rm rot}/E_{\rm grav}$ is smaller for larger cores.

\subsection{Implications for observation}

In this subsection, we will discuss the implications of our simulation results for observational studies.

We have found that when initial turbulence is strong ($\mathcal{M}=5$), regardless of the setup, $\bm{L}_{\rm core}$ does not align with the rotation axis of the clump. Therefore, in a typical star cluster-forming clump with around $\mathcal{M}=5$, it is difficult to distinguish whether the single clump is rotating or undergoing a collision using observational results of $\bm{L}_{\rm core}$. However, although it may be a rare case, if the turbulence in the clump is significantly weak, the $\bm{L}_{\rm core}$ observations may be used to distinguish between single rotating clump and colliding clump. We have confirmed that in the Rotation Setup, when turbulence is weak ($\mathcal{M}=1.5$), there is a tendency for $\bm{L}_{\rm core}$ to align with the rotation axis of the clump. For instance, if there is a velocity gradient in the observed star-forming clump and the estimated rotation axis is significantly aligned with the angular momentum of the internal dense core, this indicate that a single clump is rotating, rather than two clumps colliding. 

\begin{figure*}
\begin{center}
  \includegraphics[width=16cm]{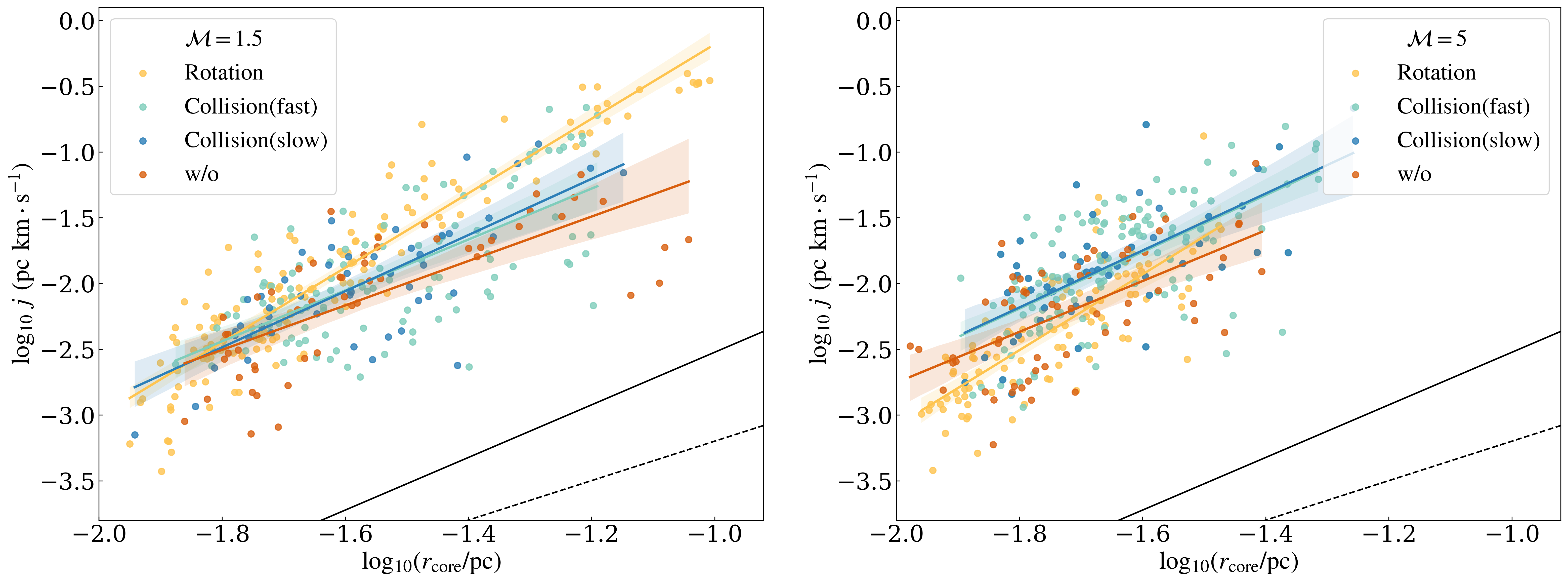}
\end{center}
\caption{Specific angular momentum, $j\equiv L_{\rm core}/M_{\rm core}$, plotted against the core radius for all $\mathcal{M}=1.5$ models (left) and $\mathcal{M}=5$ (right) models, showing the best-fit power law relation with 95\% confidence bands for each setup. The black solid and dashed lines are $j\propto r^{2}$ and $j\propto r^{1.5}$ respectively.} For $\mathcal{M}=1.5$, the slope have fits of 2.83 (Rotation), 1.94 (collision (fast)), 2.13 (collision (fast)), 1.69 (w/o). For $\mathcal{M}=5$, the slope have fits of 2.88 (Rotation), 2.16 (collision (fast)), 2.17 (collision (fast)), 1.93 (w/o). When $\mathcal{M}=1.5$, $j$ and the slope of best-fit in the Rotation Setup is relatively large. However, in $\mathcal{M}=5$, although the slope is large, $j$ in the rotation setup is not significantly different from the core in other setups. 
\label{fig:j-r_rotation}
\end{figure*}

Total specific angular momentum of cores, $j=L_{\rm core}/M_{\rm core}$, may also have important implications. Early observations have shown that $j$ are correlated with the core size $R$, following a power-law $j\propto R^{\alpha}$ \citep{1993ApJ...406..528G,Caselli_2002,2003A&A...405..639P, 2016PASJ...68...24T}. The correlation $j\propto R^{1.5}$ suggests that the rotation velocity inside the core is inherited from a turbulent cascade \citep{Chen_2018}, while $j\propto R^{2}$ is expected for solid body rotation. \citet{2018A&A...617A..27P} showed $j\propto R^{1.8-2.4}$ for the core in the L1495 ﬁlament in the Taurus molecular cloud. More recently, \citet{2023MNRAS.tmp.2141P} found $j\propto R^{1.82\pm 0.10}$ for cores in some star forming regions, suggesting that velocity gradients within cores originate from a combination of solid body rotation and turbulent motions.
Figure \ref{fig:j-r_rotation} shows the $j-R$ relation. In the Collision Setup, for both $\mathcal{M}=1.5$ and 5, $j$ is slightly higher than w/o Setup. This is due to the injection of turbulence during the collision, which increases the total kinetic energy of the core and, consequently, its specific angular momentum. However the slope index does not differ much, and the fitting curves are within the confidence intervals of w/o Setup. Slope values are close to 2, suggesting the rigid body rotation.
On the other hand, in the Rotation Setup models for weak turbulence $\mathcal{M}=1.5$, the $j-R$ relation is different from other setup models. Especially in larger cores, $j$ takes relatively large values with smaller variations, resulting in a steeper slope. As discussed earlier, in the Rotation Setup, the overall rotation of the clump is transferred to the core and has a significant impact on its angular momentum. This process of rotation inheritance may be reflected in the $j-R$ relation. 
These characteristic in $j-R$ relation could be useful for distinguishing rotating clump and colliding clumps. However, due to the large scatter from the fitting curve in the data points in Figure \ref{fig:j-r_rotation}, we can not conclusively determine the extent of the significant difference between the two setups in $j-R$ relation. Also, \citet{2023ApJ...943...76M} showed that the $j-R$ slope changes depending on the evolutionary stage of the core. Further studies considering the evolutionary stages deserve.
For strong turbulence $\mathcal{M}=5$, since the rotation of the clump is not transferred, there is no significant difference in the magnitude and variation of $j$ compared to other models. 

$\bm{B}_{\rm core}$ also reflect the environment of the clump, thus the observation of $\bm{B}_{\rm core}$ can serve as a useful clue for estimating the state of the clump. We found that in the Collision Setup, the magnetic field tends to bend globally along the shock front, which is then inherited by the core, making $\bm{B}_{\rm core}$ pairs align more easily. On the other hand, in the Rotation Setup, when $\bm{B}_{0}$ is weak, the magnetic field is easily disturbed resulting in the misalignment of $\bm{B}_{\rm core}$ pairs. In some star-forming regions, by the polarization observation, it has been suggested that magnetic fields are twisted due to the collision impact \citep{Dewangan_2018,10.1093/pasj/psaa053}. 
The magnetic field within cores formed in such regions may highly aligned along the collision interface. 
Further observations of magnetic fields within clumps and dense cores may provide useful information for diagnosing the possibility of collision and estimating clump quantities.

We have confirmed that, generally, the angle between $\bm{L}_{\rm core}$ and $\bm{B}_{\rm core}$ is random. In other words, the direction of $\bm{L}_{\rm core}$ and thus the outflow is independent of the magnetic field conditions of the clump. This fact would provide important implications for observational studies to investigate the core rotation or outflow orientation in star-forming regions \citep[e.g.,][]{Doi_2020,Xu_2022}.

To summarize this subsection, the angular momentum and mean magnetic field of dense cores reflect the environment of the clump and may be useful in distinguishing whether a star-forming clump is rotating, colliding, or neither. These parameters can be efficient tools for estimating the physical state of the star-forming clump.

\subsection{Caveats}
\label{sec:caveat}
Our analyses only focus on the early prestellar stage of the bound cores. Therefore we do not take into account the effects of stellar feedback. Feedback is critical in determining the local core-to-star efficiency and driving turbulence across various scales within clumps. Hence, feedback is expected to alter the physical core properties and the clump environment.

Also, the range of parameters investigated in this study, such as clump mass, initial gas density, magnetic field strength, turbulence intensity, collision velocity, etc., is limited, which may introduce biases in our results. However, despite these limitations, our findings provide important insights into the physical properties of bound cores, i.e., the initial conditions for star formation. 

\section{Summary}
\label{sec:Summary}
%collisionではコアは整列しない, 衝突面に沿ってalignする, 乱流が駆動されてunboundとなるコアが多い。
%回転ではコアは整列, core massが大きいと回転エネルギーの寄与が大きい, separationは大きくなる
We have investigated the properties of dense cores, including angular momentum $\bm{L}_{\rm core}$ and inner magnetic fields $\bm{B}_{\rm core}$ in the cluster-forming clump using isothermal MHD simulations with self-gravity. Two different setups were examined, including a single rotating clump and colliding clump. Our main results are summarized as follows:

\begin{itemize}
      \item[1.] In the rotating clump, for $E_{\rm clump,rot}$/$E_{\rm clump, tur}>1$ cases, $\bm{L}_{\rm core}$ inherit the rotation of parental clump. $\bm{L}_{\rm core}$ tends to align with the rotational axis of the clump, $\bm{\Omega}_{0}$,
      regardless of the initial magnetic fields strength $B_{0}$ and orientation $\theta_{0}$. However, in $E_{\rm clump,rot}$/$E_{\rm clump, tur}\sim1$ cases, there is no tendency for alignment. The turbulence intensity is an important parameter that determines whether or not the rotation of the parental clump is transferred to bound cores.
      
      \item[2.] In the colliding clump, $\bm{L}_{\rm core}$ does not inherit the rotation of the parental clump irrespective of the collision speed, turbulence strength, and initial magnetic field properties. Generally, the distribution of $\measuredangle[\bm{L}_{\rm core},\bm{\Omega}_{\rm col}]$ is found to be random. This is because, when the clumps collide, turbulence is induced, and the gas inflow into the dense cores becomes isotropic.

      \item[3.] In the colliding clump, the magnetic field tends to bend globally along the shock-compressed layer, which is then inherited by the core, making the mean magnetic field of the dense core, $\bm{B}_{\rm core}$, align with shocked layer. The collision axis is a crucial factor in determining the direction of $\bm{B}_{\rm core}$. 
      
      \item[4.] Generally, $\bm{B}_{\rm core}$ and $\bm{L}_{\rm core}$ are not aligned, and distributions of $\measuredangle[\bm{L}_{\rm core},\bm{B}_{\rm core}]$ is random. $\bm{B}_{\rm core}$ does not constrain the direction of $\bm{L}_{\rm core}$. 

      \item[5.] Regardless of the setups, the contribution of the core's rotational energy is small, accounting for $\sim 5\%$ of the gravitational energy. However, the core feature of the Rotation Setup is that the contribution of $E_{\rm rot}$ is higher than in other Setups, and the variance of $E_{\rm rot}/E_{\rm kin}$ and $E_{\rm rot}/E_{\rm grav}$ is smaller for larger cores.
      
\end{itemize}

\begin{acknowledgments}
We thank the referee for a very helpful comments. Computations described in this study were conducted using the {\tt Enzo} code \citep{2014ApJS..211...19B} and carried out on Cray XC50 at Center for Computational Astrophysics, National Astronomical Observatory of Japan. This research also made use of the yt-project a toolkit for analyzing and visualizing quantitative data \citep{2011ApJS..192....9T}. 
\end{acknowledgments}


\vspace{5mm}


\software{Enzo \citep{2014ApJS..211...19B},  
          yt \citep{2011ApJS..192....9T}
          }


\appendix

\section{Time evolution of column density}
\label{app:Time evolution of column density}

\begin{figure*}
\begin{center}
  \includegraphics[width=13.7cm]{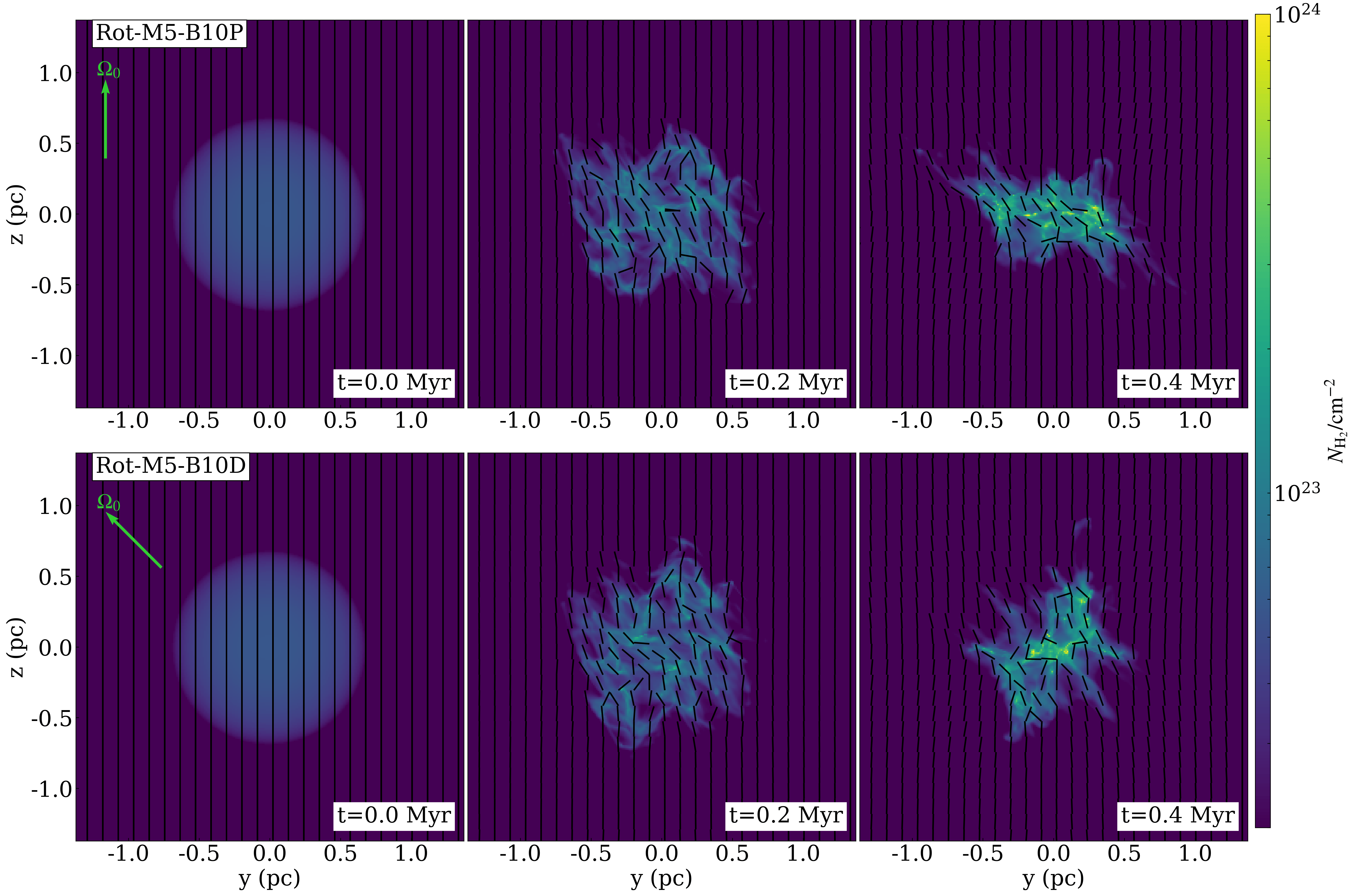}
\end{center}
\caption{Time evolution of column density along the $x$-axis for the {\tt Rot-M5-B10P} (top) and {\tt Rot-M5-B10D} (bottom) model. Snapshots at 0.0, 0.2, and 0.4 Myr are shown. Mass-weighted magnetic fields direction projected on the corresponding plane are shown as black lines. $\bm{\Omega}_{0}$ is indicated by green arrows. In both models, the gas forms a roughly perpendicular, disk-like structure with respect to the rotation axis. Other Rotation setup models also undergo similar temporal evolution, forming elongated structures that extend perpendicular to the rotation axis.}
\label{fig:rot_time_evolution}
\end{figure*}

\begin{figure*}
\begin{center}
  \includegraphics[width=13.7cm]{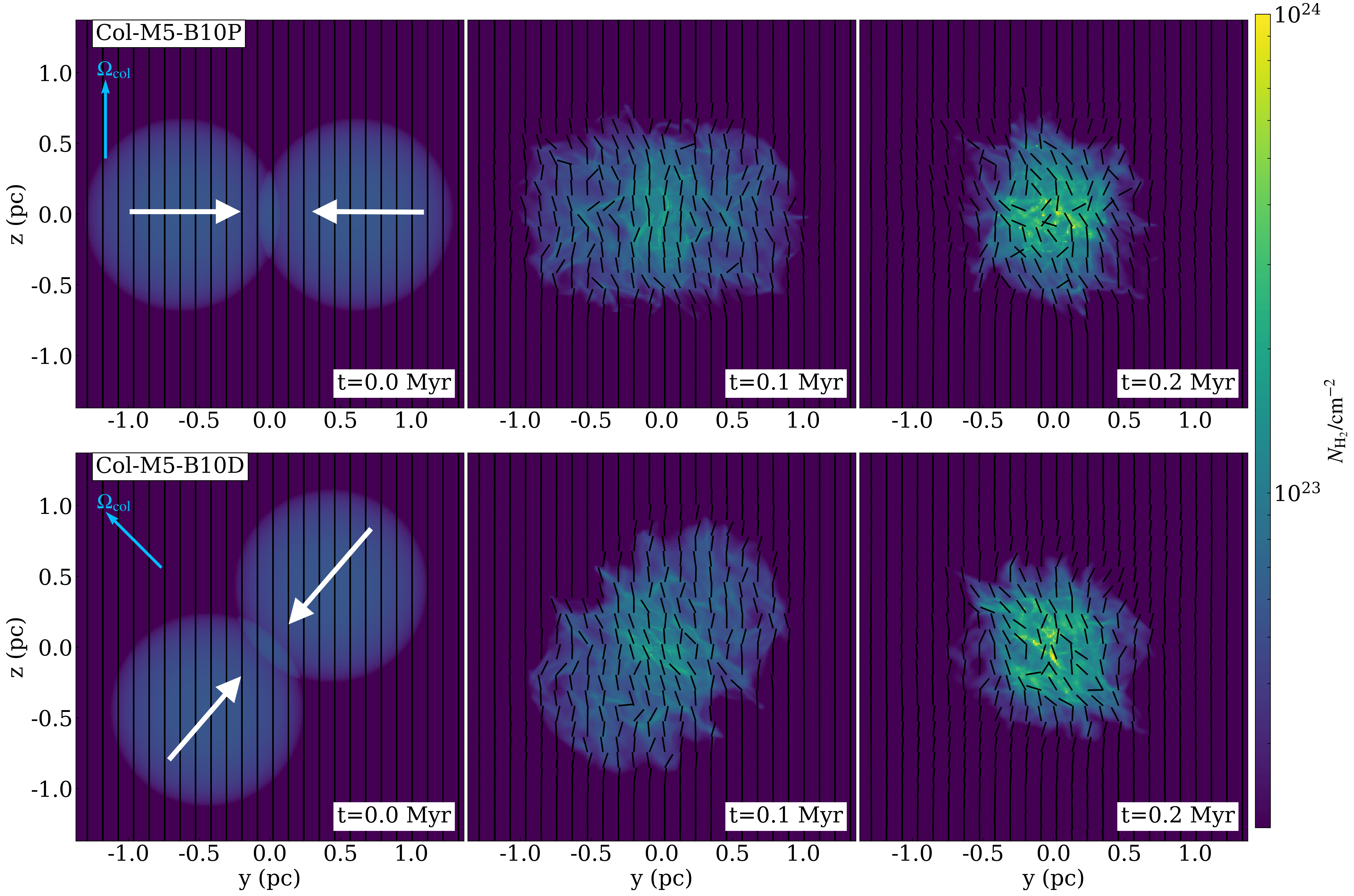}
\end{center}
\caption{Time evolution of column density along the $x$-axis for the {\tt Col-M5-B10P} (top) and {\tt Col-M5-B10D} (bottom) model. Snapshots at 0.0, 0.1, and 0.2 Myr are shown. Mass-weighted magnetic fields direction projected on the corresponding plane are shown as black lines. $\bm{\Omega}_{\rm col}$ is indicated by blue arrows. White arrows indicate the collision axis.}
\label{fig:col_time_evolution}
\end{figure*}

We show the time evolution of column density for some representative models of Rotation Setup and Collision Setup. The column density along the $x$-axis at 0.0, 0.2, and 0.4 Myr for two Rotation Setup models, {\tt Rot-M5-B10P} and {\tt Rot-M5-B10D} are shown in Figure \ref{fig:rot_time_evolution}. The initial uniform magnetic field is distorted by turbulence and clump rotation. In both models, the gas forms a roughly perpendicular, disk-like structure with respect to the rotation axis (see also Figure \ref{fig:initial_cindition}). Other Rotation setup models also undergo similar temporal evolution, forming disk-like structures that extend perpendicular to the rotation axis.

The column density along the $x$-axis at 0.0, 0.1, and 0.2 Myr for two Collision Setup models {\tt Col-M5-B10P} and {\tt Col-M5-B10D} model are shown in Figure \ref{fig:col_time_evolution}. The initial uniform magnetic field is distorted by the turbulence and clump moving. As shown in Figure \ref{fig:col_B_align}, the magnetic fields inside the clump are aligned along the shocked layer.

\section{Properties of identified bound cores}
\label{app:property}
Table \ref{tab:Properties of identified cores} gives the core properties related to the orientation parameter of $\bm{B}_{\rm core}$ and $\bm{L}_{\rm core}$, as well as the rotational energy.

\begin{deluxetable*}{lcccllc}
\tablecaption{Properties of identified bound cores}
\label{tab:Properties of identified cores}
\tablewidth{0pt}
\tablehead{
\colhead{Model name \hspace{40pt} } & \colhead{$S_{L,\Omega}$$^{\rm a}$} & \colhead{$S_{B,B_{0}}$$^{\rm b}$} & \colhead{$S_{L,B}$$^{\rm c}$} & \colhead{$\beta$~$^{\rm d}$} & \colhead{$E_{\rm rot}/E_{\rm kin}$$^{\rm e}$}  &\colhead{$N_{\rm core}$$^{\rm f}$}\\
\nocolhead{} & \nocolhead{} & \nocolhead{} & \nocolhead{} & $/10^{-2}$  & $/10^{-2}$  &\nocolhead{} 
}
\startdata
Rotation Setup              &         & &         &        &        &                                        \\ \hline
\tt{Rot-M1.5-B10P}          & 0.55    & 0.06 & 0.06    & 4.6$\,$(2.2--8.2)    & 19.3$\,$(10.0--25.9)       &    49                                 \\
\tt{Rot-M1.5-B100P}         & 0.60    & 0.44 & 0.34    & 6.3$\,$(5.1--9.6)    & 36.8$\,$(32.2--44.3)       &    24                                \\
\tt{Rot-M5-B10P}            & 0.12    & 0.22 & 0.07    & 3.2$\,$(2.2--6.1)    & 15.1$\,$(8.4--23.0)       &    32                                 \\
\tt{Rot-M5-B100P}           & 0.06    & 0.52 & 0.10    & 3.8$\,$(2.1--16.5)   & 24.7$\,$(7.1--55.3)      &    27                                \\
\tt{Rot-M1.5-B10D}          & 0.55    & -0.10 & -0.10   & 5.$\,$3(3.5--8.4)    & 18.9$\,$(13.0--30.8)      &    43                                 \\
\tt{Rot-M1.5-B100D}         & 0.37    & 0.31 & 0.08    & 7.9$\,$(5.2--13.5)    &  28.9$\,$(24.3--37.1)      &    24                              \\
\tt{Rot-M5-B10D}            & 0.19    & 0.14 & -0.13   & 7.1$\,$(4.0--11.0)    &  26.1$\,$(8.9--36.9)      &    20                               \\
\tt{Rot-M5-B100D}           & 0.08    & 0.36 & 0.05    & 7.0$\,$(2.4--11.3)   &  36.7$\,$(13.9--54.0)     &    23                               \\ \hline
w/o Setup                   &         & &         &        &        &                                          \\ \hline
\tt{w/o-M1.5-B10}           & \nodata & 0.08 & 0.06    & 1.7$\,$(0.6--3.9)    & 4.1$\,$(0.9--8.8)    &    21                                     \\
\tt{w/o-M1.5-B100}          & \nodata & 0.69 & -0.18   & 2.3$\,$(1.1--5.7)    & 10.1$\,$(3.8--19.8)      &    32                                     \\
\tt{w/o-M5-B10}             & \nodata & 0.24 & 0.06    & 7.4$\,$(2.1--13.2)    & 20.5$\,$(7.3--33.5)     &    33                                       \\
\tt{w/o-M5-B100}            & \nodata & 0.73 & 0.11    & 2.1$\,$(0.9--7.4)    & 11.1$\,$(3.4--30.6)      &    27                                     \\ \hline
Collision Setup             &         & &         &        &        &                                        \\ \hline
\tt{Col-M1.5-B10P}          & 0.08    & 0.83 & 0.09    & 0.7$\,$(0.4--2.1)    & 1.7$\,$(0.8--2.8)       &    32                                   \\
\tt{Col-M1.5-B100P}         & 0.05    & 0.88 & 0.07    & 3.6$\,$(2.2--7.0)    & 15.5$\,$(10.8--24.0)       &    29                                   \\
\tt{Col-M5-B10P}            & 0.10    & 0.41 & 0.05    & 8.8$\,$(2.3--14.0)    &  16.3$\,$(6.0--29.5)     &    22                                 \\
\tt{Col-M5-B100P}           & 0.07    & 0.67 & 0.07    & 7.3$\,$(3.2--13.5)    & 19.1$\,$(9.8--37.5)       &    62                                  \\
\tt{Col-M1.5-B10D}          & -0.20   & 0.25 & -0.20   & 1.3$\,$(0.2--3.1)    & 2.0$\,$(0.7--4.1)       &    26                                   \\
\tt{Col-M1.5-B100D}         & 0.04    & 0.39 & 0.13    & 5.5$\,$(3.1--7.6)    &  19.6$\,$(12.0--26.6)      &    23                                     \\
\tt{Col-M5-B10D}            & 0.13    & 0.18 & 0.08    & 7.0$\,$(3.3--15.4)    & 10.7$\,$(7.4--20.7)       &    30                                     \\
\tt{Col-M5-B100D}           & -0.05   & 0.46 & 0.17    & 5.2$\,$(1.8--11.7)    & 16.2$\,$(7.0--26.6)       &    59                                     \\
\tt{Col-S-M1.5-B10P}        & 0.00    & 0.59 & -0.10   & 2.3$\,$(1.1--3.3)    &  4.7$\,$(2.2--7.3)      &    24                                      \\
\tt{Col-S-M1.5-B100P}       & -0.13   & 0.68 & -0.08   & 3.3$\,$(2.2--5.7)    &  22.2$\,$(11.0--26.7)    &    21                                      \\
\tt{Col-S-M5-B10P}          & 0.03    & 0.29 & 0.03    & 4.3$\,$(2.4--7.4)    &  13.0$\,$(10.4--20.5)      &    20                                      \\
\tt{Col-S-M5-B100P}         & 0.02    & 0.66 & 0.23    & 5.9$\,$(2.7--12.1)    &  22.5$\,$(15.7--39.3)      &    50                                      \\ \hline
\enddata
\tablecomments{$^{\rm a}$ The orientation parameter $S_{L,\Omega}=(3\langle \mathrm{cos}^{2} \measuredangle[\bm{L}_{\rm core},\bm{\Omega}_{0}(\bm{\Omega}_{\rm col})]\rangle-1)/2$. 
$^{\rm c}$ The orientation parameter $S_{L,B}=(3\langle \mathrm{cos}^{2} \measuredangle[\bm{L}_{\rm core},\bm{B}_{\rm core}]\rangle-1)/2$.
$^{\rm d}$ The median and upper/lower quartiles of the rotational parameter $\beta(\equiv E_{\rm rot}/|E_{\rm grav}|)$ over all cores for each parameter set.
$^{\rm e}$ The median and upper/lower quartiles of $E_{\rm rot}/E_{\rm kin}$ over all cores for each parameter set. 
$^{\rm f}$ The total number of identified bound cores. 
}
\end{deluxetable*}

\section{Nearest neighbor core separation}
\label{app:Nearest neighbor core separation}

\begin{figure*}
\begin{center}
  \includegraphics[width=15cm]{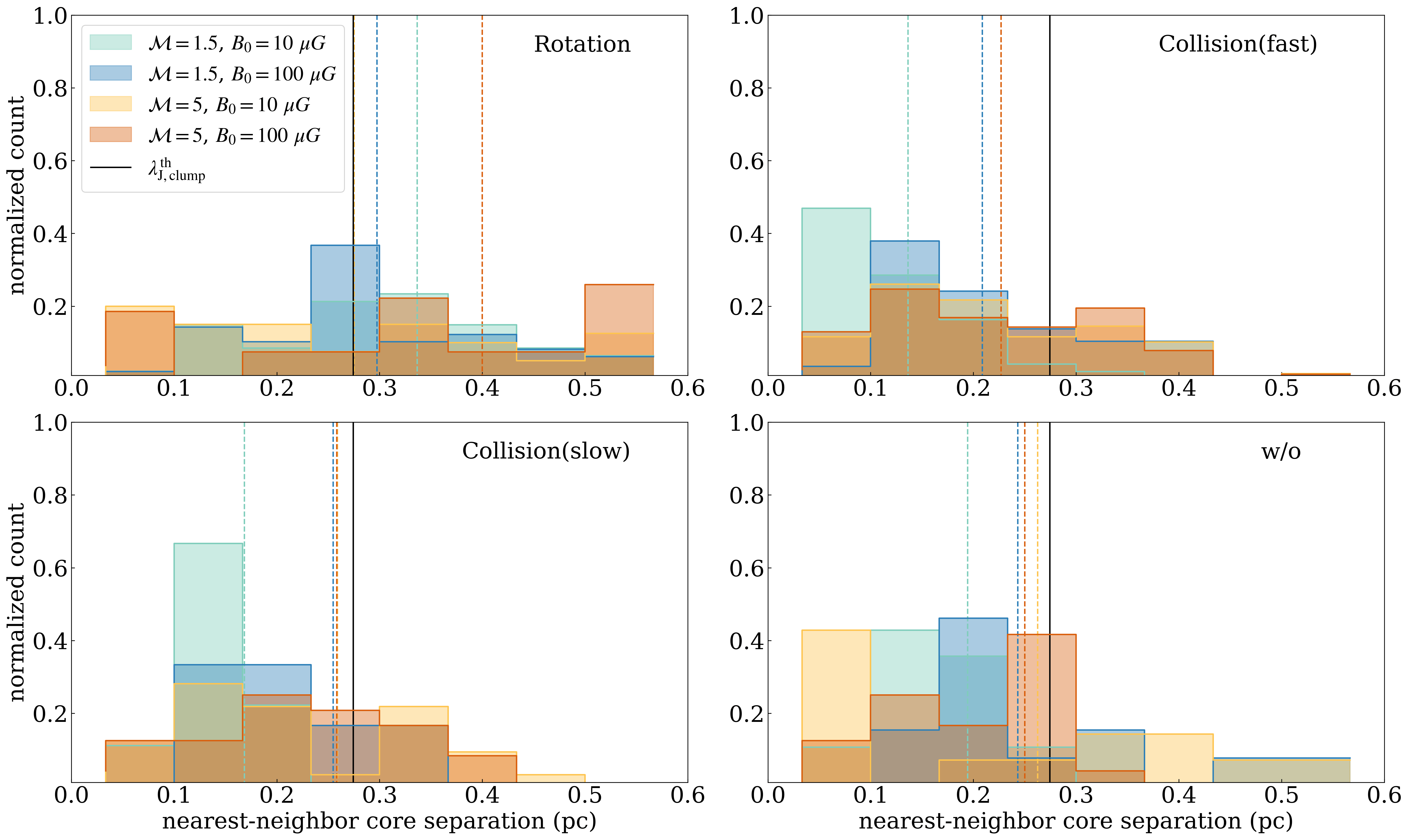}
\end{center}
\caption{Histograms of nearest neighboring separations.  The Rotation Setup models are displayed on the top left, the Collision Setup (fast) models on the top right. The models with $\theta_{0}=0^{\circ}$ and $45^{\circ}$ are presented together. The Collision Setup (slow) models are shown on the bottom left, and the w/o Setup model on the bottom right. Vertical dashed lines with different colors indicate the average separation for each parameter model. Vertical black lines represent thermal Jeans length $\lambda_{\rm J,clump}^{\rm th}$ derived using the initial condition of the clump. The Rotation Setup models have higher peak positions, means, and variances in the histogram compared to the other models, indicating that the rotational motion of the clump has an impact on the fragmentation process.
}
\label{fig:separation}
\end{figure*}

We derived the nearest neighboring separations for identified bound cores in each simulation run using the minimum spanning tree (MST) method. The MST is a graph theory technique that connects a set of points with a set of straight lines such that the total length of the lines is minimized. MST was initially introduced by \citep{10.1093/mnras/216.1.17} for astrophysical applications and has since been widely utilized in the research of the spatial distribution of star-forming objects like dense cores (e.g., \citealt{Wu_2020}; \citealt{2021A&A...646A..25Z}; Ishihara et al. 2023 in prep).

In Figure \ref{fig:separation}, we present histograms of core MST separations. The black lines indicate the thermal Jeans length $\lambda_{\rm J,clump}^{\rm th}=c_{\rm s}(\pi/G \rho_{0})^{1/2}$, where $\rho_{0}$ is the initial mass density of the clump. In the Collision Setup and w/o Setup, the peak and average values of separation distributions are comparable to or smaller than $\lambda_{\rm J,clump}^{\rm th}$. The local gas density increases due to the compression by turbulence or collisions. This density enhancement can lead to an effective Jeans length smaller than $\lambda_{\rm J,clump}^{\rm th}$, which is determined based on the initial mass density of the clump. Therefore, it is reasonable that the separation is approximately equal to or smaller than $\lambda_{\rm J,clump}^{\rm th}$. In the Rotation Setup, the variance of the distribution is larger than that in the other setups, and the averages of the separation distribution are larger, around or twice the $\lambda_{\rm J,clump}^{\rm th}$. The rotational motion of the clump is found to affect the fragmentation process of the gas and widen the scale. The centrifugal force generated by the rotational motion is presumed to have expanded the gas distribution, which has lengthened the fragmentation scale.

%\section{The correlation between $\bf{L}_{\rm core}$ and $\bf{B}_{\rm core}$ alignment and energy.}

\section[alternative title goes here]{The correlation between $\bm{L}_{\rm \lowercase{core}}$ and $\bm{B}_{\rm \lowercase{core}}$ alignment and energies of cores.}
\label{app:LB-energy}

\begin{figure*}
\begin{center}
  \includegraphics[width=13.5cm]{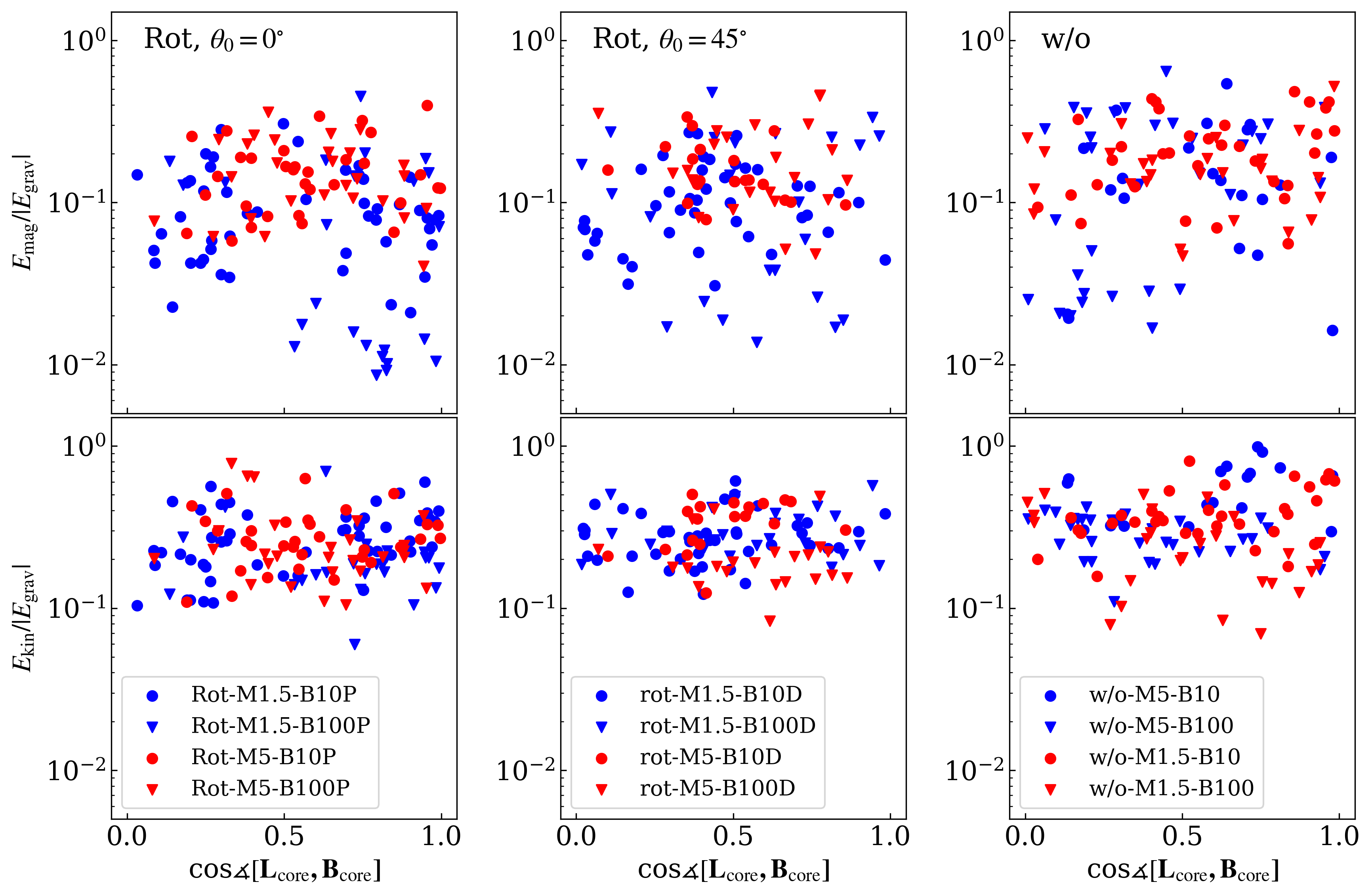}
\end{center}
\caption{Top row: Ratio between magnetic and gravitational energies, $E_{\rm mag}/|E_{\rm grav}|$, as functions of the cosine of the relative angle between $\bm{L}_{\rm core}$ and $\bm{B}_{\rm core}$, $\mathrm{cos}\,\measuredangle[\bm{L}_{\rm core},\bm{B}_{\rm core}]$. Rotation Setup models with $\theta_{0}=0^{\circ}$ are shown in the left panel, while $\theta_{0}=45^{\circ}$ models are shown in the middle panel. The right panel shows the results of w/o Setup. Bottom row: Ratio between kinetic and gravitational energies, $E_{\rm kin}/|E_{\rm grav}|$,  plotted against $\mathrm{cos}\,\measuredangle[\bm{L}_{\rm core},\bm{B}_{\rm core}]$. The independence of $\measuredangle[\bm{L}_{\rm core},\bm{B}_{\rm core}]$ with $E_{\rm mag}/|E_{\rm grav}|$ is confirmed. $\measuredangle[\bm{L}_{\rm core},\bm{B}_{\rm core}]$ is also independent of  $E_{\rm kin}/|E_{\rm grav}|$. }
\label{fig:rot_LB-ene}
\end{figure*}

\begin{figure*}
\begin{center}
  \includegraphics[width=13.5cm]{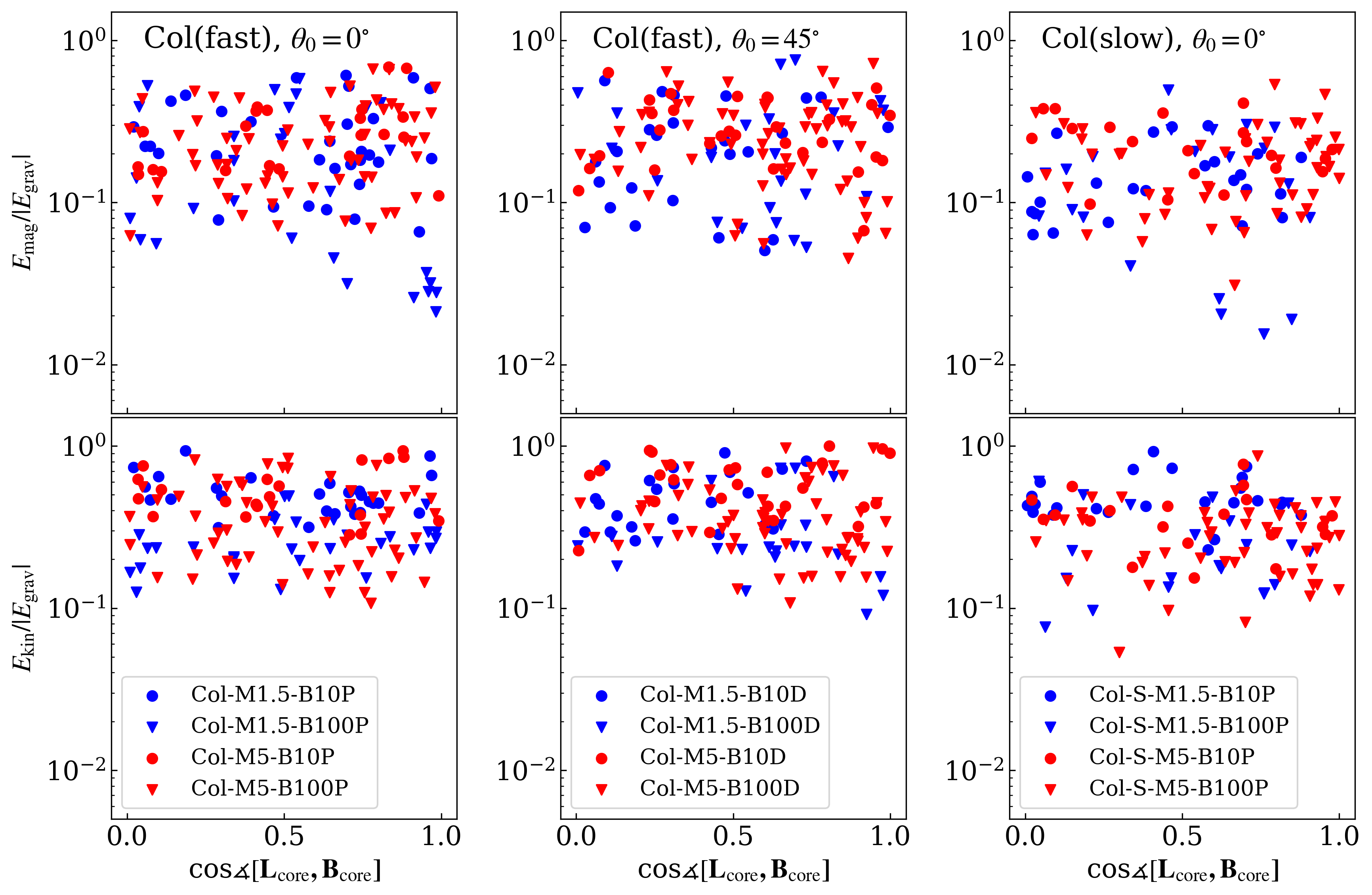}
\end{center}
\caption{Same as Figure \ref{fig:rot_LB-ene} except for Collision Setup models. Fast collision velocity cases with $\theta_{0}=0^{\circ}$ are shown in the left panel and those of $\theta_{0}=45^{\circ}$ are shown in the middle panel. Slow collision velocity cases are shown in the right panel. $\measuredangle[\bm{L}_{\rm core},\bm{B}_{\rm core}]$ does not depend on either $E_{\rm mag}/|E_{\rm grav}|$ and $E_{\rm kin}/|E_{\rm grav}|$. }
\label{fig:col_LB-ene}
\end{figure*}

Figure \ref{fig:rot_LB-ene} shows the correlation between $\measuredangle[\bm{L}_{\rm core},\bm{B}_{\rm core}]$ and energies of cores for Rotation and w/o Setup models. 
The first row of Figure \ref{fig:rot_LB-ene} presents the ratio between magnetic and gravitational energies, $E_{\rm mag}/|E_{\rm grav}|$, as functions of the cosine of the relative angle between $\bm{L}_{\rm core}$ and $\bm{B}_{\rm core}$, $\mathrm{cos}\,\measuredangle[\bm{L}_{\rm core},\bm{B}_{\rm core}]$. $\measuredangle[\bm{L}_{\rm core},\bm{B}_{\rm core}]$ is independent of $E_{\rm mag}/|E_{\rm grav}|$. 
The second row shows the ratio between kinetic and gravitational energies, $E_{\rm kin}/|E_{\rm grav}|$, as functions of $\mathrm{cos}\,\measuredangle[\bm{L}_{\rm core},\bm{B}_{\rm core}]$. $\measuredangle[\bm{L}_{\rm core},\bm{B}_{\rm core}]$ does not depend on $E_{\rm kin}/|E_{\rm grav}|$ either. 
As Figure \ref{fig:rot_LB-ene}, Figure \ref{fig:col_LB-ene} shows $E_{\rm mag}/|E_{\rm grav}|$ and $E_{\rm kin}/|E_{\rm grav}|$, as functions of $\mathrm{cos}\,\measuredangle[\bm{L}_{\rm core},\bm{B}_{\rm core}]$ for Collision Setup models.
Even in Collision Setup models, $\measuredangle[\bm{L}_{\rm core},\bm{B}_{\rm core}]$ is independent of $E_{\rm mag}/|E_{\rm grav}|$ or $E_{\rm kin}/|E_{\rm grav}|$.

\section{Energetic properties of bound and unbound cores}
\label{app:Energetic properties of cores}

Table \ref{tab:Energetic properties of cores} gives the energetic properties of cores. In this table, each value is calculated using both bound cores and unbound cores.

\begin{deluxetable}{lll}
\tablecaption{Energetic properties of bound and unbound cores}
\label{tab:Energetic properties of cores}
\tablewidth{0pt}
\tablehead{
\colhead{Model name \hspace{40pt} } & \colhead{$E_{\rm kin}/|E_{\rm grav}|\,^{\rm a}$ } &  \colhead{$E_{\rm mag}/|E_{\rm grav}|\,^{\rm b}$} \\ 
\nocolhead{} & $/10^{-2}$  & $/10^{-2}$ } 

\startdata
Rotation Setup              &                    &                 \\ \hline
\tt{Rot-M1.5-B10P}          &   26.1$\,$(18.5--35.9)  & 10.5$\,$(6.2--23.9)                                  \\
\tt{Rot-M1.5-B100P}         &   18.9$\,$(14.9--24.7)  & 18.3$\,$(1.8--55.8)                                  \\
\tt{Rot-M5-B10P}            &   29.7$\,$(21.5--36.5)  & 20.0$\,$(12.3--32.0)                                  \\
\tt{Rot-M5-B100P}           &   23.4$\,$(16.7--45.4)  & 47.1$\,$(20.5--94.4)                               \\
\tt{Rot-M1.5-B10D}          &   26.2$\,$(21.0--32.3)  & 13.8$\,$(8.4--26.6)    \\
\tt{Rot-M1.5-B100D}         &   29.2$\,$(24.3--38.4)  & 23.3$\,$(3.8--47.9)                                   \\
\tt{Rot-M5-B10D}            &   34.3$\,$(23.1--45.6)  & 27.3$\,$(14.1-46.5)   \\
\tt{Rot-M5-B100D}           &   22.9$\,$(17.1--36.0)  & 60.0$\,$(29.0-95.8)                                   \\ \hline
w/o Setup                   &                    &                    \\ \hline
\tt{w/o-M1.5-B10}           &   51.28$\,$(39.6--65.8)  & 30.9$\,$(13.8--99.1)                                      \\
\tt{w/o-M1.5-B100}          &   26.9$\,$(20.8--35.3)  & 30.4$\,$(11.2--96.3)                                         \\
\tt{w/o-M5-B10}             &   38.7$\,$(29.7--53.4)  & 39.3$\,$(22.6--60.5)                                          \\
\tt{w/o-M5-B100}            &   25.5$\,$(18.0--40.6)  & 64.8$\,$(18.3--141.9)                                         \\ \hline
Collision Setup             &                    &                   \\ \hline
\tt{Col-M1.5-B10P}          &   107.6$\,$(65.1--248.5)  & 168.6$\,$(61.1--351.3)    \\
\tt{Col-M1.5-B100P}         &   29.6$\,$(23.4--48.0)    & 85.1$\,$(18.1--340.1)                     \\
\tt{Col-M5-B10P}            &   135.0$\,$(84.7--227.3)  & 135.0$\,$(67.3--281.9)                                  \\
\tt{Col-M5-B100P}           &   56.3$\,$(31.7--92.1)    & 135.0$\,$(52.1--286.1)                \\
\tt{Col-M1.5-B10D}          &   143.5$\,$(76.1--287.5)    & 130.8$\,$(55.0--266.1)                                      \\
\tt{Col-M1.5-B100D}         &   46.9$\,$(26.1--83.4)    & 146.0$\,$(42.2--390.1)                                          \\
\tt{Col-M5-B10D}            &   101.9$\,$(68.5--184.1)    & 110.1$\,$(64.2--218.6)                                        \\
\tt{Col-M5-B100D}           &   58.0$\,$(32.5--100.5)    & 117.5$\,$(55.1--243.1)                                         \\
\tt{Col-S-M1.5-B10P}        &   50.7$\,$(41.1--72.7)    & 60.2$\,$(14.5--121.0)                                        \\
\tt{Col-S-M1.5-B100P}       &   23.9$\,$(16.0--28.5)    & 17.8$\,$(8.9--60.8)                                         \\
\tt{Col-S-M5-B10P}          &   60.1$\,$(39.0--88.6)    & 72.3$\,$(38.4--154.6)                                           \\
\tt{Col-S-M5-B100P}         &   34.5$\,$(20.4--53.5)    & 67.1$\,$(23.1--158.9)                                          \\ \hline
\enddata
\tablecomments{The values in this table are calculated including both bound and unbound cores.
        $^{\rm a}$ The median and upper/lower quartiles of $E_{\rm kin}/|E_{\rm grav}|$ for each parameter set.
        $^{\rm b}$ The median and upper/lower quartiles of $E_{\rm mag}/|E_{\rm grav}|$ for each parameter set.
}
\end{deluxetable}
\clearpage{}

\bibliography{sample631}{}
\bibliographystyle{aasjournal}


%% Include this line if you are using the \edit1, \replaced, \deleted

\end{document}